%% file: main.tex
\documentclass{aa}

\usepackage{graphicx}
\usepackage{amsmath,amsfonts,amssymb}
\usepackage{esvect}
\usepackage{txfonts}
\usepackage{color}
\usepackage{xcolor}
\usepackage{natbib}
\usepackage{subcaption}
\usepackage{float}
\usepackage{multirow}
\usepackage{dblfloatfix}
\usepackage{afterpage}
\usepackage{ifthen}
\usepackage[morefloats=12]{morefloats}
\usepackage{placeins}
\usepackage{multicol}
\usepackage{siunitx}
\usepackage{multirow}
\DeclareSIUnit\parsec{pc}
\DeclareSIUnit\lightyear{ly}
\DeclareSIUnit\Kcmb{K_{cmb}}
\DeclareSIUnit\year{yr}
\bibpunct{(}{)}{;}{a}{}{,}
\usepackage[switch]{lineno}
\definecolor{linkcolor}{rgb}{0.6,0,0}
\definecolor{citecolor}{rgb}{0,0,0.75}
\definecolor{urlcolor}{rgb}{0.12,0.46,0.7}
\usepackage[breaklinks, colorlinks, urlcolor=urlcolor,
    linkcolor=linkcolor,citecolor=citecolor,pdfencoding=auto]{hyperref}
\hypersetup{linktocpage}

\input{Planck}

\def\fsky{f_{\textrm{sky}}}
\def\WMAP{\textit{WMAP}}

\def\K {\textrm{K}}
\def\Ka{\textrm{Ka}}
\def\Q {\textrm{Q}}
\def\V {\textrm{V}}
\def\W {\textrm{W}}

\newcommand{\BP}{\textsc{BeyondPlanck}}
\newcommand{\cosmoglobe}{\textsc{Cosmoglobe}}

\newcommand{\npipe}[0]{\texttt{NPIPE}}

\newcommand{\planck}[0]{\textit{Planck}}

\let\vec\vv

\def\inv{^{-1}}

\begin{document}

\title{\textsc{Cosmoglobe} DR1 results. II. Constraints on isotropic cosmic birefringence from reprocessed \WMAP\ and \Planck\ LFI data}

\input{authors.tex}
\authorrunning{Eskilt et al.}
\titlerunning{\cosmoglobe\ Cosmic Birefringence}

\abstract{Cosmic birefringence is a parity-violating effect that might have rotated the plane of linearly polarized light of the cosmic microwave background (CMB) by an angle $\beta$ since its emission. This has recently been measured to be non-zero at a statistical significance of $3.6\sigma$ in the official \textit{Planck} PR4 and 9-year \WMAP\ data. In this work, we constrain $\beta$ using the reprocessed \BP\ LFI and \cosmoglobe\ DR1 \WMAP\ polarization maps. These novel maps have both lower systematic residuals and a more complete error description than the corresponding official products. Foreground $EB$ correlations could bias measurements of $\beta$, and while thermal dust $EB$ emission has been argued to be statistically non-zero, no evidence for synchrotron $EB$ power has been reported. Unlike the dust-dominated \Planck\ HFI maps, the majority of the LFI and \WMAP\ polarization maps are instead dominated by synchrotron emission. Simultaneously constraining $\beta$ and the polarization miscalibration angle, $\alpha$, of each channel, we find a best-fit value of $\beta=0.35^{\circ}\pm0.70^{\circ}$ with LFI and \WMAP\ data only. When including the \Planck\ HFI PR4 maps, but fitting $\beta$ separately for dust-dominated, $\beta_{>70\,\mathrm{GHz}}$, and synchrotron-dominated channels, $\beta_{\leq 70\,\mathrm{GHz}}$, we find $\beta_{\leq 70\,\mathrm{GHz}}=0.53^{\circ}\pm0.28^\circ$. This differs from zero with a statistical significance of $1.9\sigma$, and the main contribution to this value comes from the LFI 70\,GHz channel. While the statistical significances of these results are low on their own, the measurement derived from the LFI and \WMAP\ synchrotron-dominated maps agrees with the previously reported HFI-dominated constraints, despite the very different astrophysical and instrumental systematics involved in all these experiments.
}


\maketitle

%\hypersetup{linkcolor=black}
%\tableofcontents
%\hypersetup{linkcolor=red}
%\clearpage

\section{Introduction}
\label{sec:introduction}

The standard model of cosmology, $\Lambda$CDM, predicts no parity-violating physics on cosmological scales. There are however extensions of the standard model that allow it, one example being an ultra-light axion-like field $\phi$ that couples to the electromagnetic field tensor $F_{\mu \nu}$,
\begin{equation}
\mathcal{L} \supset - \frac14 g_{\gamma \phi} \phi F_{\mu \nu} \Tilde{F}^{\mu \nu},
\end{equation}
where $\mathcal{L}$ is the Lagrangian density, $g_{\gamma \phi}$ is a coupling constant and $\Tilde{F}^{\mu \nu}$ is the Hodge dual of $F_{\mu\nu}$ (see for instance a recent review by \citealp{Komatsu:2022nvu}). This term effectively causes a rotation of the plane of linear polarization of electromagnetic waves as they propagate through empty space \citep{Carroll:1989vb, Carroll:1991zs, Harari:1992ea}. The predicted rotation angle $\beta$ as a function of the direction of the line of sight $\hat{\mathbf n}$ is
\begin{equation}
\beta(\hat{\mathbf n})
    =\frac12 g_{\gamma \phi}\left[\phi(\eta_{\textrm{o}})-\phi(\eta_{\textrm{e}},r\hat{\mathbf n})\right],
\end{equation}
where $\eta_{\textrm{o}}$ and $\eta_{\textrm{e}}$ are the conformal times at observation and emission of the photons, respectively, and $r\equiv\eta_{\textrm{o}} - \eta_{\textrm{e}}$ is the conformal distance from the observer to the emitter.

The cosmic microwave background (CMB) is the oldest polarized light in the universe, and therefore among our most sensitive probes of a light axion-like field. Raw CMB polarization measurements are typically provided in terms of the Stokes parameters, $Q$ and $U$, that measure two orthogonal modes of linear polarization in a given coordinate system. When using such measurements to constrain cosmic birefringence, it is convenient to transform these into coordinate-independent $E$ and $B$ modes following \citet{Kamionkowski:1996zd} and \citet{Seljak:1996gy}. In this representation, cosmic birefringence rotates the intrinsic $E$ and $B$-modes emitted at the last scattering surface into the $E^{\textrm{o}}$ and $B^{\textrm{o}}$ modes we observe today \citep{Lue:1998mq}. Assuming that the cosmic birefringence angle is isotropic and adopting the convention that a positive $\beta$ defines a clockwise rotation of the plane of linear polarization on the sky, these rotations are given as 
\begin{equation}
\begin{bmatrix}
E_{\ell m}^{\textrm{o}}\\
B_{\ell m}^{\textrm{o}}
\end{bmatrix} = \begin{bmatrix}
\cos(2\beta) & -\sin(2\beta) \\
\sin(2\beta) & \cos(2\beta)
\end{bmatrix}\begin{bmatrix}
E_{\ell m}\\
B_{\ell m}
\end{bmatrix}.
\end{equation}
Defining the angular power spectrum as $C^{XY}_\ell = \frac{1}{2\ell+1}\sum_m X_{\ell m}Y^*_{\ell m}$, one can show that \citep{Feng:2004mq}
\begin{equation}
    \label{eq:simple_cb}
    C^{EB,\, \textrm{o}}_\ell = \frac{\tan(4\beta)}{2}\left(C^{EE,\, \textrm{o}}_\ell - C^{BB,\, \textrm{o}}_\ell\right) + \frac{C^{EB}_\ell}{\cos(4\beta)}.
\end{equation}
The latter term is the intrinsic $EB$ power spectrum of the CMB, which is predicted to be zero in $\Lambda$CDM, and we accordingly neglect it in this work.

No instrument is perfect. All polarization-sensitive CMB detectors have an intrinsic miscalibration angle $\alpha$, that represents the difference between the true polarization angle of the detector and that assumed in the analysis. This angle induces a spurious rotation of the plane of the polarized electromagnetic waves and is therefore for a single frequency channel fully degenerate with $\beta$. To account for this instrumental systematic in an estimate of cosmic birefringence, one needs to replace $\beta \rightarrow \alpha + \beta$ in Eq.~\eqref{eq:simple_cb}. The calibration uncertainty in $\alpha$ will then induce a corresponding systematic uncertainty in $\beta$.

A novel method to break the $\alpha+\beta$ degeneracy was introduced by \citet{Minami:2019ruj} and \citet{MinamiKomatsu:2020}, who proposed to exploit the differing multipole behaviour between CMB and polarized foreground emission to self-calibrate the miscalibration angles. In this approach, an important assumption is that cosmic birefringence would have a negligible impact on the photons originating from our Galaxy while only substantially impacting the CMB photons which have travelled much farther. \citet{Minami:2020odp} applied this method to the \planck\ Public Release 3 (PR3) polarization data, and found a best-fit value of $\beta = 0.35^\circ \pm 0.14^\circ$, which nominally disfavors $\beta=0^\circ$ at a statistical significance of $2.4\sigma$.

This analysis on the HFI channels was further improved by \citet{Diego-Palazuelos:2022dsq}, who applied the same method to the \planck\ Public Release 4 (PR4) data, often called ``\npipe'' \citep{planck2020-LVII}. This data set has a higher signal-to-noise ratio than PR3 due to the inclusion of more data, and it has lower systematic uncertainties. The updated analysis yielded a birefringence angle constraint of $\beta = 0.30^\circ \pm 0.11^\circ$, corresponding to a 21\,\% smaller uncertainty than the original analysis.

Neither of these measurements explicitly accounted for the intrinsic $EB$ correlation in polarized thermal dust emission, as it has not been directly measured. \planck\ did, however, detect parity-odd dust $TB$ correlations \citep{planck2016-l11A}. The origin of the measured $TB$ correlations has been hypothesized to be due to a misalignment between local magnetic fields and dust filaments \citep{Huffenberger:2019mjx, Clark:2021kze}, which would also give rise to an intrinsic $EB$ power spectrum of dust emission.

Using an ansatz inspired by the filamentary model of \citet{Clark:2021kze}, \citet{Diego-Palazuelos:2022dsq} found that the statistical significance of the cosmic birefringence increased to $3.3\sigma$ with $\beta = 0.36^\circ \pm 0.11^\circ$. However, they also noted that this model slightly affects the value of $\beta$ as a function of sky fraction.

In a follow-up work, one of the authors of this work included the rest of the \planck\ polarization data by incorporating the \planck\ Low Frequency Instrument (LFI) bands of \planck\ Public Release 4 \citep{Eskilt:2022wav}. These frequencies are dominated by synchrotron emission, for which no evidence of an intrinsic $EB$ correlation has been claimed to date \citep{Martire:2021gbc, Rubino-Martin:2023fya}. A positive measurement of $\beta$ using synchrotron-dominated maps alone could suggest that $C^{EB,\, \mathrm{dust}}_{\ell}$ is not the cause of the recent non-zero measurements of isotropic cosmic birefringence. Indeed, even without including the filamentary dust $EB$ model for the HFI channels, the inclusion of the LFI maps alone increased the statistical significance of a non-zero cosmic birefringence angle to $3.3\sigma$ at $\beta = 0.33^\circ \pm 0.10^\circ$.

\cite{Eskilt:2022wav} also analyzed the frequency behaviour of the birefringence signature, and found that it is consistent with being frequency-independent, which is precisely the prediction of an ultra-light axion that couples to electromagnetism. At the same time, an explanation based on Faraday rotation of magnetic fields is disfavored because of its strong frequency dependency, $\beta \propto 1/\nu^2$, where $\nu$ is frequency. Recently, \citet{Eskilt:2023nxm} showed that the signal is not consistent with early dark energy coupling to photons.

An analysis of possible systematic effects that could bias measurements of $\beta$ in the HFI channels was performed by \cite{Diego-Palazuelos:2022cnh}. The authors showed that the positive measurement of $\beta$ is robust against beam leakage, intensity-to-polarization leakage, and cross-polarization effects. However, they did warn that foreground $EB$ could potentially bias a birefringence measurement, as also reported by \cite{Diego-Palazuelos:2022dsq}.

Finally, \cite{Eskilt:2022cff} also included the 9-year \WMAP\ measurements between 23 and 94\,GHz in the cosmic birefringence analysis. Even though the \WMAP\ observations have a low signal-to-noise ratio with respect to $\beta$ by themselves, cross-correlating these channels with LFI and HFI channels provides slightly smaller error bars on $\beta$. Specifically, the baseline result yielded $\beta = 0.342_{\phantom{\circ}-0.091^{\circ}}^{\circ + 0.094^\circ}$ using nearly full-sky data, which is a $3.6\sigma$ measurement of a non-zero birefringence angle. They also found a consistent result, $\beta = 0.37^\circ \pm 0.14^\circ$, with a much larger mask for which thermal dust $EB$ correlations have been argued to be mostly positive \citep{Clark:2021kze, Diego-Palazuelos:2022dsq}.

As of today, there are no other useful public CMB satellite polarization data available to include in this particular analysis framework, and even the amount of ground-based or sub-orbital data is limited. The reason for this is that the estimator developed by \citet{Minami:2019ruj} and \citet{MinamiKomatsu:2020} requires a sufficient amount of polarized foreground emission to break the degeneracy between $\alpha$ and $\beta$, but ground-based telescopes tend to point their instruments away from foreground-dominated parts of the sky to maximize their CMB sensitivity. In addition, none of the public ground-based telescope data can guarantee a miscalibration angle that is small compared to $\beta \sim 0.3^\circ$. Therefore, we expect little to be gained from including these data sets as they are now. See, however, \citet{Cornelison:2022zrc} for recent attempts to calibrate the BICEP3 instrument to achieve a precision of $<0.1^{\circ}$.

While the amount of new data is currently limited, systematic uncertainties from both astrophysical foregrounds and instrumentation remain an important issue regarding these cosmic birefringence measurements. In that respect, we note that there has been a major effort in the community aiming to build a single coherent end-to-end CMB Bayesian analysis framework that simultaneously accounts for instrumental, astrophysical, and cosmological parameters. This work originally started within the \planck\ collaboration \citep{planck2014-a12} as an effort to establish a coherent model of the astrophysical sky through Bayesian analysis. This method was later generalized to also account for instrumental effects by the \BP\ collaboration \citep[][and references therein]{bp01}, who also used this novel framework to derive \Planck\ LFI sky maps with lower systematic residuals than the official products \citep{bp10}.

\cosmoglobe\footnote{\url{https://cosmoglobe.uio.no}} is a global Open Science initiative that aims to simultaneously apply this framework to all available state-of-the-art data sets, and build one coherent model of the radio, microwave, and sub-millimeter sky based on all these observations. The first major \cosmoglobe\ data release (called DR1) is described in a suite of four papers, and includes the first joint end-to-end processing of the \WMAP\ and \Planck\ LFI data sets. The main results, as defined by frequency maps and preliminary astrophysical and cosmological results, are summarized by \citet{Watts:2023vdc}; perhaps the most striking outcome from this work is a set of \WMAP\ polarization maps with significantly lower large-scale instrumental systematics than the official products. 

This paper focuses on birefringence measurements from the reprocessed \cosmoglobe\ \WMAP\ and \BP\ LFI maps. Not only are these maps cleaner in terms of instrumental effects than previous products, but most of these particular frequencies are also dominated by polarized synchrotron emission. A constraint on cosmic birefringence derived from \WMAP\ and LFI alone is therefore intrinsically interesting because it will be associated with very different astrophysical uncertainties than the previous results from HFI and thermal dust-dominated maps. A third advantage of the new products is that they are associated with a large ensemble of posterior-based samples, each corresponding to a different realization of instrumental systematic uncertainties. This novel product allows for a much more complete uncertainty estimation than previous traditional pipelines. In particular, the \WMAP\ maps produced by \citet{Watts:2023vdc} significantly improve the statistical treatment of poorly constrained transmission imbalance modes, and they allow for a direct unbiased estimate of the polarized sky without explicit post-processing.

The rest of the paper is organized as follows. Section~\ref{sec:bp} describes the \BP\ and \cosmoglobe\ data sets, while Sect.~\ref{sec:method} gives a brief review of the cosmic birefringence methodology. We present our main results in Sect.~\ref{sec:results}, and we draw conclusions in Sect.~\ref{sec:conclusions}.

\section{\BP\ and \cosmoglobe\ data}
\label{sec:bp}

In this paper, we focus primarily on the reprocessed \cosmoglobe\ DR1 \WMAP\ \citep{Watts:2023vdc}, \BP\ LFI \citep{bp01}, and \Planck\ PR4 HFI \citep{planck2020-LVII} polarization sky maps, although we also analyze some combinations of the legacy 9-year \WMAP\ \citep{Bennett:2013} and \Planck\ PR4 LFI \citep{planck2020-LVII} data. The \BP\ and \cosmoglobe\ products are derived using the end-to-end Gibbs sampler implemented in the \texttt{Commander3}\footnote{\url{https://github.com/Cosmoglobe/Commander/}} codebase \citep{bp03}. This method draws samples for each band's calibration and noise parameters while conditioning on a sky model. Then the sky model is sampled while conditioning on the same calibration and noise parameters. This iterative process maps out the joint posterior distribution, accounting for correlations between instrumental and sky parameters in a statistically coherent framework \citep[see, e.g.,][for further discussion of the Gibbs sampling methodology and philosophy]{geman:1984,bp04}.

As discussed by \cite{bp10}, the reprocessed LFI maps resulting from this procedure provide better control of gain uncertainty than the PR3 and PR4 analyses. For these reasons, we consider the \BP\ \texttt{Commander3} approach to yield the most accurate and best characterized \Planck\ LFI maps to date. In particular, the 44\,GHz map, which was plagued by null test failures even after the PR4 analysis, is now of sufficient quality to be used for cosmological analyses. For example, \citet{bp11} and \citet{bp12} used these maps to make robust measurements of the large-scale polarized CMB and reionization optical depth $\tau$. 

The \BP\ project was a pathfinder that focused specifically on \Planck\ LFI. Now \cosmoglobe\ aims to apply this unified framework to jointly analyze as many state-of-the-art data sets as possible. The first major application to a new data set is \WMAP, as first described by \citet{bp17}. The \WMAP\ experiment \citep{Bennett:2013} observed the microwave sky with ten polarization-sensitive differencing assemblies (DAs) at multiple frequencies, including \K-band (23\,GHz), \Ka-band (33 GHz), \Q-band (41 GHz), \V-band (61 GHz), and \W-band (94 GHz). As presented in a companion paper by \citet{Watts:2023vdc}, a joint analysis of \WMAP{} and \Planck{} LFI time-ordered data within the \cosmoglobe\ framework has now yielded maps that are essentially free of poorly measured transmission imbalance modes, resulting in the best consistency between \Planck{} LFI and \WMAP{} to date. 

\cosmoglobe\ DR1 provides updated maps for both \WMAP\ and LFI, and in principle, we could therefore have used also the \cosmoglobe\ DR1 LFI maps in the current birefringence analysis. However, as noted by \citet{Minami:2020odp} and \citet{Diego-Palazuelos:2022dsq}, it is generally advantageous to exploit only cross-correlation spectra in order to maximize the overall signal-to-noise ratio of the estimator in question. Since the \BP\ data release provides half-mission split maps for all LFI channels, while \cosmoglobe\ DR1 currently only provides co-added full-mission maps \citep{bp10}, we use the \BP\ half-mission LFI maps in the following. Furthermore, \citet{Watts:2023vdc} show that the \cosmoglobe\ LFI maps are very similar to the \BP\ LFI maps, differing only by 1--2\,\muK\ on large angular scales. Specifically, for LFI we use 200 half-mission \BP\ Gibbs samples for each channel, while for \WMAP\ we use the first 200 samples of the first \cosmoglobe\ DR1 main chain\footnote{Available at \url{https://cosmoglobe.uio.no}.} \citep{Watts:2023vdc}. This Gibbs chain only sampled the \V-band and \W-band maps once every two and four Gibbs iterations, respectively, due to the high cost of mapmaking for these bands, resulting in a total of 100 \V-band samples and 50 \W-band samples to be processed in the following.

For \Planck\ HFI, we use the latest official \Planck\ PR4 products for all frequency channels. In terms of ancillary data, we use the official beam profiles provided by the \WMAP\ and \Planck\ collaborations \citep{Bennett:2013,planck2014-a05}.

\section{Method}
\label{sec:method}

In this section, we provide a brief review of the methodology introduced and developed by \citet{Minami:2019ruj}, \citet{MinamiKomatsu:2020} and \citet{Diego-Palazuelos:2022dsq}. As noted above, the fundamental assumption of \citet{Minami:2019ruj} is that cosmic birefringence would have a negligible impact on polarized foreground emission from our own Galaxy, and only significantly impacts the CMB photons which have traveled for almost 14 billion years. The plane of linear polarization of the foreground emission is therefore rotated only by the miscalibration angle $\alpha$, while the CMB photons are rotated by $\alpha+\beta$. We can therefore write the observed $E$ and $B$-modes as,
\begin{align}
\begin{bmatrix}
E_{\ell m}^{\textrm{o}}\\
B_{\ell m}^{\textrm{o}}
\end{bmatrix}\nonumber &= \begin{bmatrix}
\cos(2\alpha) & -\sin(2\alpha) \\
\sin(2\alpha) & \cos(2\alpha)
\end{bmatrix}\begin{bmatrix}
E^{\textrm{fg}}_{\ell m}\\
B^{\textrm{fg}}_{\ell m}
\end{bmatrix}\\
&+\begin{bmatrix}
\cos(2\alpha+2\beta) & -\sin(2\alpha+2\beta) \\
\sin(2\alpha+2\beta) & \cos(2\alpha+2\beta)
\end{bmatrix}\begin{bmatrix}
E^{\textrm{CMB}}_{\ell m}\\
B^{\textrm{CMB}}_{\ell m}
\end{bmatrix} + \begin{bmatrix}
E^{\textrm{n}}_{\ell m}\\
B^{\textrm{n}}_{\ell m}
\end{bmatrix},
\label{eq:obsEB}
\end{align}
where ``fg'', ``CMB'', and ``n'' denote the foreground, CMB, and noise components, respectively.

We define the ensemble-averaged power spectrum as $\langle C^{XY}_{\ell}\rangle = \delta_{\ell \ell'}\delta_{mm'} \langle X_{\ell m}Y^{*}_{\ell' m'}\rangle$. When computing this for the observed $E$ and $B$ modes defined in Eq.~\eqref{eq:obsEB}, one finds \citep{Minami:2019ruj}
\begin{align}
    \nonumber
    \langle C_\ell^{EB,\, \textrm{o}}\rangle &= \frac{\tan(4\alpha)}{2}\left(\langle C_\ell^{EE,\, \textrm{o}}\rangle - \langle C_\ell^{BB,\, \textrm{o}}\rangle \right)\\
    \nonumber
    &+ \frac{\sin(4\beta)}{2\cos(4\alpha)}\left(\langle C_\ell^{EE,\, \text{CMB}}\rangle - \langle C_\ell^{BB, \,\text{CMB}}\rangle \right)\\
    \label{eq:singleinstrumentbeta}
    &+ \frac{1}{\cos(4\alpha)}\langle C_\ell^{EB,\, \text{fg}}\rangle  + \frac{\cos(4\beta)}{\cos(4\alpha)}\langle C_\ell^{EB,\, \text{CMB}}\rangle.
\end{align}
This equation allows us to break the degeneracy between $\alpha$ and $\beta$ by using the polarized foreground emission to calibrate $\alpha$. However, $\alpha$ might be biased if we do not include the intrinsic $EB$ correlations of the foreground emission. $C^{EB,\, \mathrm{fg}}_{\ell}$ has not been directly measured for synchrotron or dust, but there is substantial indirect evidence that it is non-zero at least for dust \citep{Clark:2021kze, Diego-Palazuelos:2022dsq, Vacher:2022mvr}.

This paper will focus on synchrotron-dominated channels, in part because there has been no detection of a non-zero $EB$ correlation for this component \citep{Martire:2021gbc, Rubino-Martin:2023fya}. In the following, we will therefore assume that $C^{EB,\, \textrm{fg}}_{\ell} = 0$ for synchrotron channels, while for dust-dominated channels, we adopt the filamentary thermal dust model for estimating the $EB$ correlation as explained below.

$\Lambda$CDM predicts no parity-violation at the last scattering surface, hence, we set $C^{EB,\, \textrm{CMB}}_\ell = 0$. One could wonder if the recent measurements of cosmic birefringence could be explained by this term. However, \citet{Fujita:2022qlk} conclude that primordial chiral gravitational waves, which give rise to intrinsic $EB$ correlations of the CMB, can not cause a cosmic birefringence measurement of $\beta \sim 0.3^\circ$ due to an overproduction of $BB$ power relative to observational constraints.

By extending the analysis to multiple frequency channels, labelled $i$ and $j$, respectively, one can show that \citep{MinamiKomatsu:2020}
\begin{align}
    \nonumber
    \begin{bmatrix}
      \langle C_\ell^{E_iE_j, \,\textrm{o}}\rangle\\
        \langle C_\ell^{B_iB_j,\, \textrm{o}}\rangle
    \end{bmatrix} &= \mathbf{R}^{\alpha}_{ij} \begin{bmatrix}
        \langle C_\ell^{E_iE_j,\, \text{fg}}\rangle\\
          \langle C_\ell^{B_iB_j,\, \text{fg}}\rangle
    \end{bmatrix}+\mathbf{D}^{\alpha}_{ij}\begin{bmatrix}
        \langle C_\ell^{E_iB_j,\, \text{fg}}\rangle\\
          \langle C_\ell^{B_iE_j,\, \text{fg}}\rangle
    \end{bmatrix}&\\
    \label{eq:c_ee_c_bb}
    &+\mathbf{R}^{\alpha+\beta}_{ij}\begin{bmatrix}
        \langle C_\ell^{E_iE_j,\, \text{CMB}}\rangle\\
          \langle C_\ell^{B_iB_j, \,\text{CMB}}\rangle
    \end{bmatrix}+\delta_{ij} \begin{bmatrix}
    \langle C_\ell^{E_iE_j, \,\text{n}}\rangle\\
      \langle C_\ell^{B_iB_j,\, \text{n}}\rangle
    \end{bmatrix},
\end{align}
and
\begin{align}
        \nonumber
      \langle C_\ell^{E_iB_j,\, \textrm{o}}\rangle &= \left(R^{\alpha}_{ij}\right)^T\begin{bmatrix}
        \langle C_\ell^{E_iE_j,\, \text{fg}}\rangle\\
          \langle C_\ell^{B_iB_j,\, \text{fg}}\rangle
    \end{bmatrix}+\left(D^{\alpha}_{ij}\right)^T\begin{bmatrix}
        \langle C_\ell^{E_iB_j,\, \text{fg}}\rangle\\
          \langle C_\ell^{B_iE_j,\, \text{fg}}\rangle
    \end{bmatrix}\\
    \label{eq:c_eb}
    &+\left(R^{\alpha+\beta}_{ij}\right)^T \begin{bmatrix}
        \langle C_\ell^{E_iE_j, \,\text{CMB}}\rangle\\
          \langle C_\ell^{B_iB_j,\, \text{CMB}}\rangle
    \end{bmatrix},
\end{align}
where
\begin{align}
    \mathbf{R}^{\theta}_{ij} &= \begin{bmatrix}
        \cos(2\theta_i)\cos(2\theta_j) & \sin(2\theta_i)\sin(2\theta_j) \\
        \sin(2\theta_i)\sin(2\theta_j) & \cos(2\theta_i)\cos(2\theta_j)
    \end{bmatrix},\\
    \mathbf{D}^{\theta}_{ij}&= \begin{bmatrix}
      -\cos(2\theta_i)\sin(2\theta_j) & -\sin(2\theta_i)\cos(2\theta_j) \\
        \phantom{-}\sin(2\theta_i)\cos(2\theta_j) & \phantom{-}\cos(2\theta_i)\sin(2\theta_j)
    \end{bmatrix},\\
    R^{\theta}_{ij} &= \begin{bmatrix}
        \phantom{-}\cos(2\theta_i)\sin(2\theta_j)\\
        -\sin(2\theta_i)\cos(2\theta_j)
    \end{bmatrix},\\
    D^{\theta}_{ij} &= \begin{bmatrix}
        \phantom{-}\cos(2\theta_i)\cos(2\theta_j)\\
        -\sin(2\theta_i)\sin(2\theta_j)
    \end{bmatrix}.
\end{align}
The different frequencies in these expressions may be actually different frequency channels from either the same or different experiments, or they may be different splits of the same frequency channel, for instance half-mission maps \citep[e.g.,][]{planck2016-l02}. In the latter case, we assume that both split maps have the same miscalibration angle $\alpha_i$, while in the former case, we assume that they are different.

In addition, we allow $\beta$ to depend on data sets, and we denote the corresponding birefringence angle $\beta_i$. This allows us to derive independent constraints on cosmic birefringence by specific frequency channels, while still being able to exploit information in complementary channels to constrain foreground contributions and the miscalibration angles. Thus, a given result will be characterized by an overall analysis configuration that defines which data sets are included in the overall analysis, as well as a specification of which channels are used to constrain $\beta$. 

Equations~\eqref{eq:c_ee_c_bb} and \eqref{eq:c_eb} can be combined into one equation as follows,
\begin{align}
\nonumber
\langle C_\ell^{E_iB_j,\, \textrm{o}}\rangle =& \left(R^{\alpha}_{ij}\right)^T\left(\textbf{R}^{\alpha}_{ij}\right)^{-1} \begin{bmatrix}
      \langle C_\ell^{E_iE_j,\, \textrm{o}}\rangle\\
        \langle C_\ell^{B_iB_j,\, \textrm{o}}\rangle
    \end{bmatrix}+\\
     \label{eq:full_equation}
    &\bigg[\left(R^{\alpha+\beta}_{ij}\right)^T-\left(R^{\alpha}_{ij}\right)^T\left(\textbf{R}^{\alpha}_{ij}\right)^{-1}\textbf{R}^{\alpha+\beta}_{ij} \bigg]\begin{bmatrix}
        \langle C_\ell^{E_iE_j,\, \text{CMB}}\rangle\\
          \langle C_\ell^{B_iB_j,\, \text{CMB}}\rangle
    \end{bmatrix},
\end{align}
where, for now, we have set the foreground $EB$ correlation to zero.
We combine the equations for all combinations of $(i, j)$
into vectors, and define both a vector of observed spectra,
$\vec{C}^{\mathrm o}_\ell = \begin{bmatrix}
  C^{E_i E_j,\, \mathrm o}_\ell, C^{B_i B_j,\, \mathrm o}_\ell, C^{E_i B_j,\, \mathrm o}_\ell
\end{bmatrix}^T$ as well as a vector of beam-smoothed theoretical $\Lambda \text{CDM}$ spectra, $\vec{C}^{\Lambda \text{CDM}}_\ell = \begin{bmatrix}
  C^{E_i E_j, \,\Lambda \text{CDM}}_\ell, C^{B_i B_j, \,\Lambda \text{CDM}}_\ell
\end{bmatrix}^T$. In all these expressions, we exclude intra-channel auto-correlations $(i, i)$ in order to avoid noise biases.

We now want to constrain $\alpha_i$ and $\beta_i$ using Eq.~\eqref{eq:full_equation} for different combinations of $(i, j)$. In practice, we reorganize these equations into a vector $\vec{v}^T_{\ell}$, which now may be written in the following matrix form,
\begin{equation}
    \label{eq:v_vector}
    \vec{v}^T_{\ell} \equiv \textbf{A}\vec{C}^{\mathrm{o}}_\ell - \textbf{B}\vec{C}^{\Lambda \text{CDM}}_\ell= 0,
\end{equation}
where
\begin{align}
    \textbf{A}_{ij} = &\begin{bmatrix}- \left(R^{\alpha}_{ij}\right)^T\left(\textbf{R}^{\alpha}_{ij}\right)^{-1}, 1\end{bmatrix},\\
    \nonumber
    \textbf{B}_{ij} = &\bigg[\left(R^{\alpha+\beta}_{ij}\right)^T-\left(R^{\alpha}_{ij}\right)^T
    \left(\textbf{R}^{\alpha}_{ij}\right)^{-1}\textbf{R}^{\alpha+\beta}_{ij}\bigg].
\end{align}
This vector has an associated covariance matrix that reads \citep{Minami:2019ruj, Minami:2020xfg, MinamiKomatsu:2020}
\begin{align}
    \label{eq:M_cov}
    \textbf{M}_\ell &= \textbf{A}\text{Cov}(\vec{C}^\mathrm{o}_\ell, \vec{C}^\mathrm{o}_\ell)\textbf{A}^T,
\end{align}
where we take the covariance of the observed spectra to be
\begin{equation}
    \label{eq:single-multipole-cov}
    \text{Cov}(C^{XY,\,\textrm{o}}_\ell, C^{ZW,\,\textrm{o}}_\ell) = \frac{C^{XZ,\,\textrm{o}}_\ell C^{YW,\,\textrm{o}}_\ell +  C^{XW,\,\textrm{o}}_\ell C^{YZ,\,\textrm{o}}_\ell}{(2\ell+1)f_{\textrm{sky}}},
\end{equation}
and $f_{\textrm{sky}}$ is the fraction of sky used for the analysis. As in previous works, we avoid including terms of the form $C^{EB,\, \mathrm{o}}_{\ell}C^{XY,\, \mathrm{o}}_{\ell}$ on the right-hand side of Eq.~\eqref{eq:single-multipole-cov} due to the large statistical fluctuations of $C^{EB,\, \mathrm{o}}_{\ell}$ \citep{MinamiKomatsu:2020}.

As discussed by \citet{Minami:2019ruj} and \citet{MinamiKomatsu:2020}, it is reasonable to assume $\vec{v}^T_{\ell}$ to be approximately Gaussian distributed, and one may then define the following likelihood,
\begin{equation}
    \ln \mathcal{L} = -\frac12 \left[ \sum_{b} \vec{v}^T_{b} \textbf{M}^{-1}_b \vec{v}_{b} + \ln \textbf{M}_b \right],
\end{equation}
where $\vec{v}^T_{b}$ depends on $\alpha_i$ and $\beta_i$ while $\textbf{M}_b$ only depends on $\alpha_i$. We additionally bin both the vector and covariance matrix over $\Delta \ell$ to reduce statistical fluctuations,
\begin{align}
    \vec{v}_{b} &= \frac{1}{\Delta \ell} \sum_{\ell \in b} \vec{v}_{\ell},\\
    \textbf{M}_b &= \frac{1}{\Delta \ell ^2}\sum_{\ell \in b} \textbf{M}_\ell.
\end{align}
To actually derive the posterior distribution for these parameters with either one or multiple channel-dependent $\beta$, we use the publicly available Markov chain Monte Carlo sampler \texttt{emcee} \citep{ForemanMackey:2012ig}, coupled to this likelihood. We adopt uniform priors on both $\alpha_i$ and $\beta_i$ in the following.

The observed power spectra are computed using \texttt{PolSpice}\footnote{\url{http://www2.iap.fr/users/hivon/software/PolSpice/}} \citep{Chon:2003gx}. We apply a mask similar to previous work \citep{Diego-Palazuelos:2022dsq, Eskilt:2022cff} which removes pixels in which the intensity of the carbon-monoxide (CO) line is stronger than $45\,\mathrm{ K_{RJ}\, km\, s^{-1}}$. We also exclude polarized point sources using the official \planck{} point source masks. We define the effective sky fraction coverage as \citep{Hivon:2002, Challinor:2004pr}
\begin{equation}
    \fsky = \frac{1}{N_{\text{pix}}}\frac{\left(\sum^{N_{\text{pix}}}_{i=1}w_i^2\right)^2}{ \sum^{N_{\text{pix}}}_{i=1}w_i^4}.
\end{equation}
Here, $N_\text{pix}$ is the number of pixels and $w_i$ is the weight of the apodized mask at pixel $i$.

Unlike previous LFI-based analyses which used the official \Planck\ half-mission maps for 30 and 44\,GHz and detector split maps for 70\,GHz \citep{planck2020-LVII, Eskilt:2022wav, Eskilt:2022cff}, we use \BP\ half-mission maps for all channels in this paper, since no \BP\ detector split maps are available at this time \citep{bp01}. The second \Planck\ half-mission split map has a few unobserved regions of the sky, which are masked. Because of this, we find that the sky fraction drops from $f_{\textrm{sky}} = 0.92$, as used in previous work \citep{Diego-Palazuelos:2022dsq, Eskilt:2022cff}, to $f_{\textrm{sky}} = 0.90$.

We use \texttt{CAMB}\footnote{\url{https://github.com/cmbant/CAMB}} \citep{Lewis:2000} to compute the theoretical $\Lambda$CDM power spectra using the cosmological parameters from \planck\ 2018 \citep{planck2016-l06}, and we smooth these spectra with the \Planck\ PR4 and 9-year \WMAP\ beam transfer function matrices $b^{X}_\ell$, respectively, together with the HEALPix\footnote{\url{http://healpix.jpl.nasa.gov}} \citep{Gorski:2004by} pixel window functions $w_{\text{pix}, \ell}$, such that
\begin{equation}
    \vec{C}^{\Lambda \text{CDM}}_\ell = \begin{bmatrix}
    C^{EE,\,\mathtt{CAMB}}_{\ell}  b^{E_i}_\ell b^{E_j}_\ell w^i_{\text{pix}, \ell}w^j_{\text{pix}, \ell} \\C^{BB,\,\mathtt{CAMB}}_{\ell} b^{B_i}_\ell b^{B_j}_\ell w^i_{\text{pix}, \ell}w^j_{\text{pix}, \ell}
    \end{bmatrix}.
\end{equation}

To include the effect of $EB$ correlations from thermal dust emission, we adopt the ansatz proposed by \cite{Diego-Palazuelos:2022dsq}, which was inspired by the work of \cite{Clark:2021kze}. Specifically, we relate the dust $EB$ power spectrum to the \planck{} measured dust $TB$ and $TE$ correlations as follows \citep{planck2016-l11A},
\begin{equation}
\label{eq:dust_eb}
C^{E_iB_j,\, \textrm{dust}}_\ell = A_\ell C^{E_iE_j,\, \textrm{dust}}_{\ell} \sin(4\psi_\ell).
\end{equation}
Here $A_\ell \geq 0$ is represented as four free parameters in the multipole ranges $51 \leq \ell \leq 130$, $131 \leq \ell \leq 210$, $211 \leq \ell \leq 510$, and $511 \leq \ell \leq 1490$ with flat positive priors, while $\psi_\ell$ is the misalignment angle between local magnetic fields and the long axis of filamentary structures averaged over some multipole range. The misalignment angle is given by \citep{Clark:2021kze}
\begin{equation}
    \psi_\ell = \frac12 \arctan \frac{C^{TB,\, \textrm{dust}}_\ell}{C^{TE,\, \textrm{dust}}_\ell}.
\end{equation}
We find $\psi_\ell$ by smoothing the observed $TB$ and $TE$ spectra from the \planck\ PR4 353\,GHz channel for the sky mask used in the analysis. We apply $\psi_\ell$ from 353\,GHz channel to all dust-dominated maps, namely the HFI channels and the \W-band from \WMAP.
To incorporate this model into the analysis, we follow \citet{Diego-Palazuelos:2022dsq} and \citet{Eskilt:2022cff} by defining the matrices
\begin{align}
    \mathbf{A}_{\ell, ij} = &\left[- \left(\Lambda^{\alpha}_{\ell, i j}\right)^T \left(\mathbf{\Lambda}^{\alpha}_{\ell, i j}\right)^{-1},\, 1\right],\\
    \nonumber
    \textbf{B}_{\ell, ij} =&\bigg[\left(R^{\alpha+\beta}_{ij}\right)^T-\left(\Lambda^{\alpha}_{\ell, i j}\right)^T\left(\mathbf{\Lambda}^{\alpha}_{\ell, i j}\right)^{-1}\textbf{R}^{\alpha+\beta}_{ij}\bigg]\,,
\end{align}
where
\begin{align}
\mathbf{\Lambda}^{\alpha}_{\ell, i j} &= \mathbf{R}^{\alpha}_{ij} + \mathbf{D}^{\alpha}_{ij}\mathbf{F}_\ell,\\
    \left(\Lambda^{\alpha}_{\ell, i j}\right)^T &= \left(R^{\alpha}_{ij}\right)^T + \left(D^{\alpha}_{ij}\right)^T\mathbf{F}_\ell,\\
    \mathbf{F}_\ell &= A_\ell \sin(4\psi_\ell)\begin{bmatrix}
        1 & 0\\
        1 & 0
    \end{bmatrix}\,.
\end{align}
These modified matrices $\mathbf{A}_\ell$ and $\mathbf{B}_\ell$ replace the $\mathbf{A}$ and $\mathbf{B}$ matrices in Eqs.~\eqref{eq:v_vector} and \eqref{eq:M_cov}. Note that we do not apply this dust $EB$ model to cross-correlations that include at least one synchrotron-dominated map. More specifically, we make the assumption that ${C^{E_{\mathrm{dust}}B_{\mathrm{synch}}}_{\ell}=C^{E_{\mathrm{synch}}B_{\mathrm{dust}}}_{\ell}=C^{E_{\mathrm{synch}}B_{\mathrm{synch}}}_{\ell}=0}$.

\begin{figure}
\centering
\includegraphics[width=\linewidth]{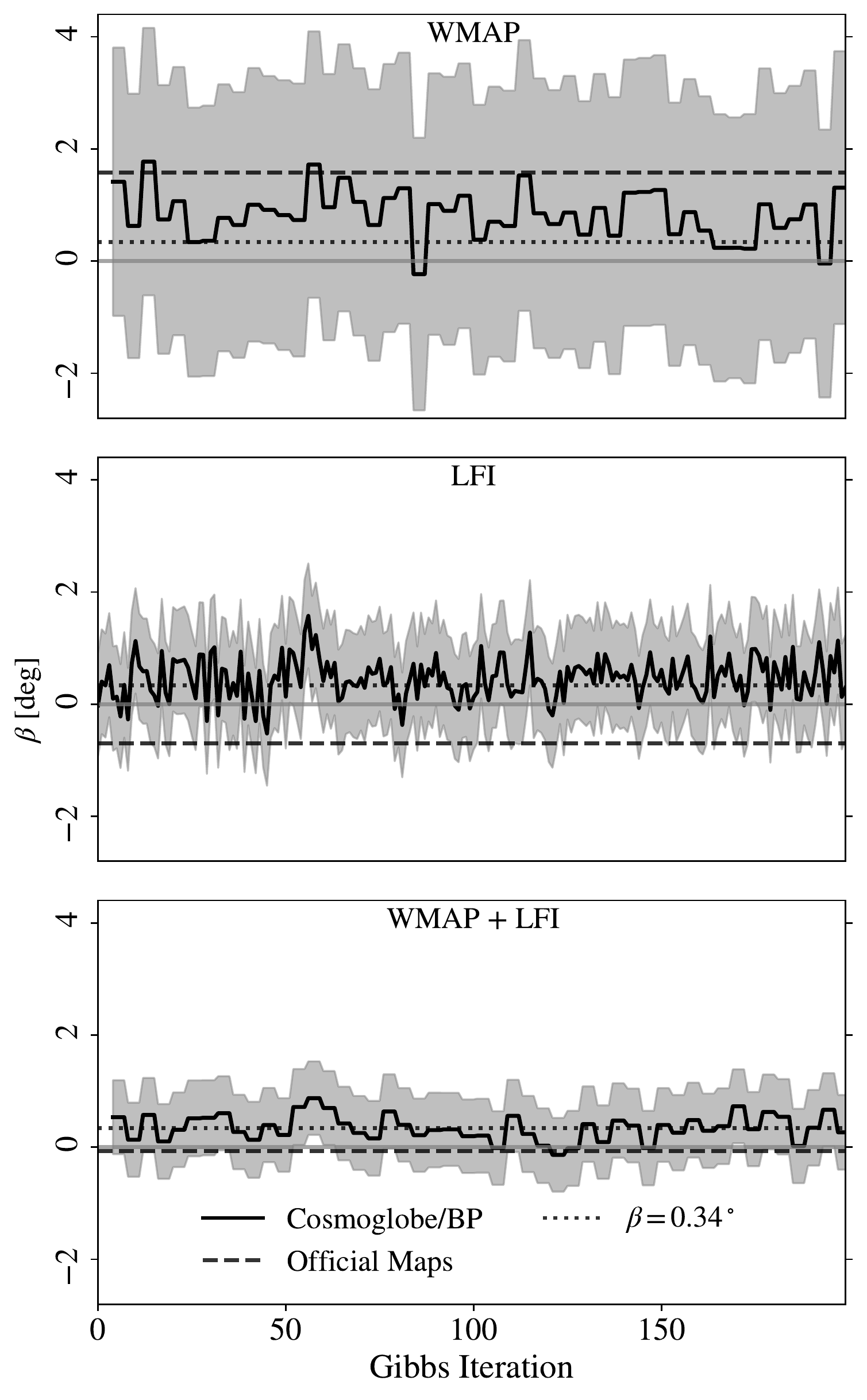}
\caption{Measurements of cosmic birefringence over Gibbs iterations. The black line shows $\beta$ with a grey $1\sigma$ band. The dashed lines show the central value of the measurement made with the corresponding 9-year \WMAP\ or \Planck\ PR4 maps, which we find to have the same white noise uncertainty as a single \cosmoglobe\ or \BP\ realization.}
\label{fig:cb-joint}
\end{figure}

\begin{table}[t]       %  Table 1
  \begingroup                                                                            
  \newdimen\tblskip \tblskip=5pt
  \caption{Summary of posterior mean cosmic birefringence angles, $\beta$, as derived for various data combinations. The first column indicates all data sets included in a given analysis, while the second column indicates the frequencies that are actually used to constrain $\beta$. The third column lists the values derived in this paper using the reprocessed \Planck\ LFI and \WMAP\ data, and the fourth column lists the corresponding values derived from the official products. All uncertainties indicate 68\,\% errors; for results derived from the reprocessed products, these include both statistical and systematic contributions, while for the results derived from the official products, they only include statistical contributions.
    \label{tab:beta}}
  \nointerlineskip                                                                                                                                                                                     
  \vskip -8mm
  \footnotesize                                                                                                                                      
  \setbox\tablebox=\vbox{ %                                                                                                                                                                             
  \newdimen\digitwidth                                                                                                                          
  \setbox0=\hbox{\rm 0}
  \digitwidth=\wd0
  \catcode`*=\active
  \def*{\kern\digitwidth}
  \newdimen\signwidth
  \setbox0=\hbox{+}
  \signwidth=\wd0
  \catcode`!=\active
  \def!{\kern\signwidth}
  \newdimen\decimalwidth
  \setbox0=\hbox{.}
  \decimalwidth=\wd0
  \catcode`@=\active
  \def@{\kern\signwidth}
  \halign{ \hbox to 1.2in{#\leaderfil}\tabskip=0.2em&
    \hfil#\hfil\tabskip=0.2em&
    \hfil$#$\hfil\tabskip=0.2em&        
    \hfil$#$\hfil\tabskip=0em\cr
  \noalign{\doubleline}
  \omit{Analysis configuration}\hfil&\omit{Fit channels}\hfil&\hfil \beta_{\mathrm{reproc}}\, \mathrm{[deg]} \hfil&\hfil \beta_{\mathrm{official}}\, \mathrm{[deg]} \hfil\cr
\omit{}\hfil&\omit{}\hfil& \hfil f_{\mathrm{sky}}=0.90 \hfil&\hfil f_{\mathrm{sky}}=0.92 \hfil\cr
  \noalign{\vskip 3pt\hrule\vskip 5pt}
  \WMAP& All & 0.81\pm2.43 & 1.58\pm2.40\cr
  \noalign{\vskip 2pt}     
  LFI& All & 0.47\pm1.00 & -0.70\pm0.95!\cr
  \noalign{\vskip 2pt}       
  HFI& All & \cdots & 0.36\pm0.11\cr
  \noalign{\vskip 2pt}     
  LFI+\WMAP& All & 0.35\pm0.70 & -0.07\pm0.64!\cr
  \noalign{\vskip 2pt}     
  LFI+HFI& 30 & -0.07\pm0.75! & \cdots\cr
  \omit& 44 &  0.10\pm0.74 & \cdots\cr
  \omit& 70 &  0.85\pm0.44 & \cdots\cr  
  \omit& LFI & 0.52\pm0.37 & \cdots\cr
  \omit& HFI & 0.22\pm0.11 & \cdots\cr
  \omit& All & 0.24\pm0.11 & 0.39\pm0.10\cr
  \noalign{\vskip 2pt}       
  LFI+HFI+\WMAP& \WMAP & -0.12\pm0.46! & \cdots\cr
  \omit& LFI & 0.53\pm0.31 & \cdots\cr
  \omit& $\le$\,70\,GHz & 0.53\pm0.28 & \cdots\cr
  \omit& >\,70\,GHz & 0.23\pm0.10 & \cdots\cr  
  \omit& HFI & 0.26\pm0.10 & \cdots\cr  
  \omit& All & 0.26\pm0.10 & 0.34\pm0.09\cr  
    \noalign{\vskip 3pt\hrule\vskip 5pt}   
  }}
  \endPlancktablewide                                                                                                                                            
  \endgroup
\end{table}

To derive the final constraints, we modify the publicly available code\footnote{\url{https://github.com/LilleJohs/Cosmic_Birefringence}} presented by \cite{Eskilt:2022cff} to include the \BP\ and \cosmoglobe\ channels rather than the 9-year \WMAP\ and LFI PR4 maps. Although the main focus of this paper is to measure cosmic birefringence from synchrotron-dominated maps, we also perform a joint analysis with the HFI channels which are all dust-dominated, and we also include the \WMAP\ \W-band, which with the new \cosmoglobe\ DR1 processing appears usable for cosmological analysis \citep{Watts:2023vdc}.

Following previous analyses \citep{Minami:2020odp, Diego-Palazuelos:2022dsq, Eskilt:2022wav, Eskilt:2022cff}, we use the multipole range $\ell_{\textrm{min}} \leq \ell \leq \ell_{\textrm{max}}$, where  $\ell_{\textrm{min}} = 51$ and $\Delta \ell = 20$. For all of LFI and \WMAP-based analyses, we use $\ell_{\textrm{max}} = 990$, and when also including HFI data, we extend the upper range to $\ell_{\textrm{max}} = 1490$.

We apply this machinery to each individual sample in the \BP\ and \cosmoglobe\ frequency map ensembles, and not the corresponding posterior mean maps. This allows us to propagate both systematic and statistical uncertainties to the final error budgets for $\alpha$ and $\beta$. We find the distribution of $\beta$ to be closely approximated by a Gaussian distribution, and we therefore report the combined uncertainty of $\beta$ over Gibbs samples by adding the two terms in quadrature, i.e., $\sigma_\beta = \sqrt{\sigma_{\textrm{syst}}^2 +\sigma_{\textrm{stat}}^2}$, where $\sigma_{\textrm{syst}}^2$ is the variance measured between Gibbs samples and $\sigma_{\textrm{stat}}^2$ is the white noise uncertainty for a single realization.

\section{Results}
\label{sec:results}

\subsection{\WMAP+LFI analysis}

\begin{figure}
\centering
\includegraphics[width=\linewidth]{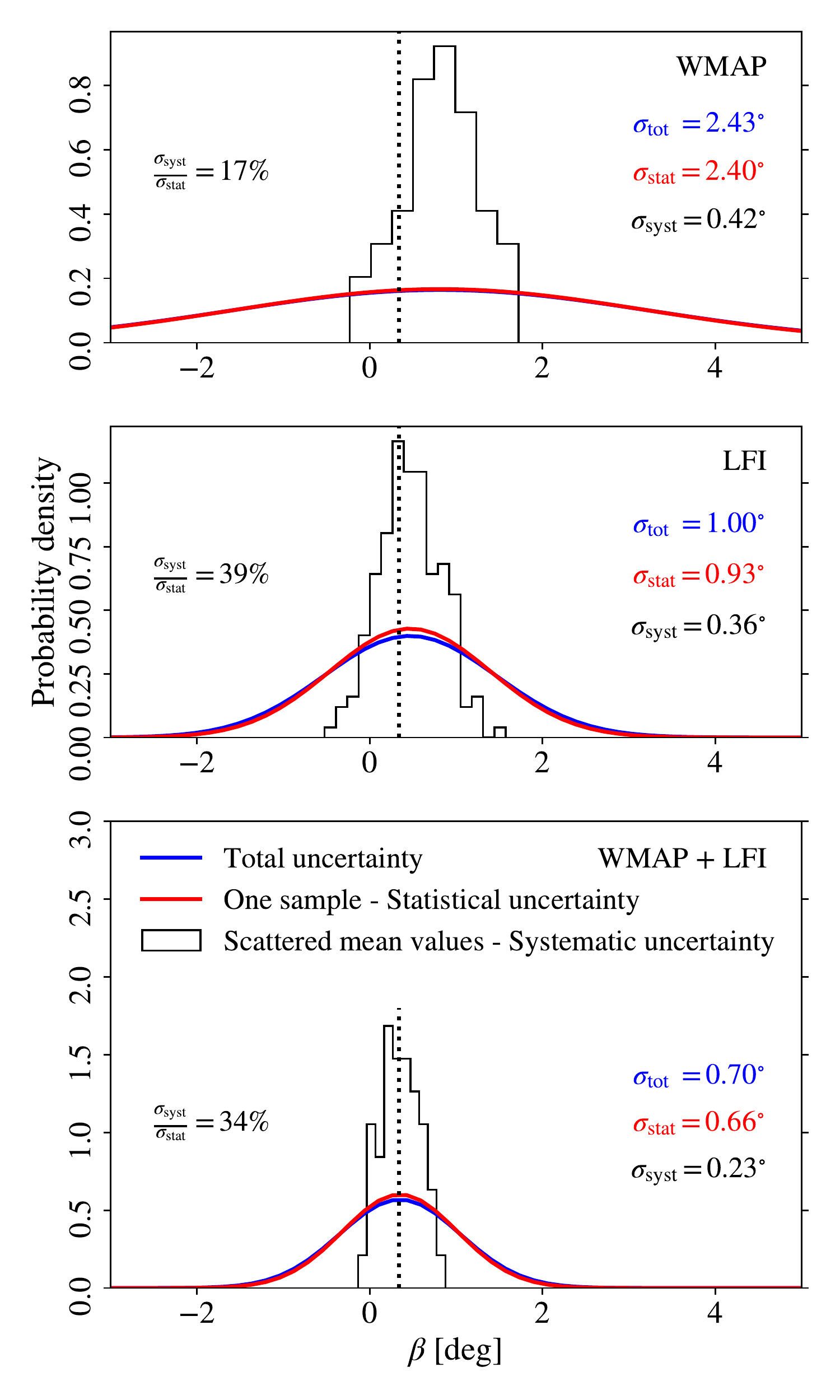}
\caption{Probability density of the measurements from Fig.~\ref{fig:cb-joint}. The black histograms show the scatter in the central values and the red lines show the statistical uncertainties. The blue lines show the total uncertainties. The black dotted line shows $\beta = 0.34^\circ$.}
\label{fig:wmap_lfi_joint_hist}
\end{figure}

We first present results derived from \Planck\ LFI and \WMAP\ only, both as processed by the \BP\ and \cosmoglobe\ framework. We begin with the case of fitting one global birefringence angle, $\beta$, across all frequencies. Markov chain traceplots for this case are shown in Fig.~\ref{fig:cb-joint}, in which panels from top to bottom show \WMAP-only, LFI-only, and \WMAP-plus-LFI results. The scatter in the central values from sample to sample illustrates the systematic uncertainties due to variations in instrumental and astrophysical parameters, while the statistical error bars (indicated as grey $1\sigma$ bands) are due to uncorrelated instrumental noise. We find these to be indistinguishable from sample to sample. Since the \W{} channel is only processed every fourth main Gibbs iteration in the \cosmoglobe\ DR1 data set \citep{Watts:2023vdc}, we also perform this measurement only every fourth Gibbs iteration for \WMAP. Overall, we see in this figure that the systematic uncertainties account for a relatively small fraction of the total scatter for \WMAP, primarily due to the higher white noise level, while they are more important for the LFI and combined analyses, which have higher signal-to-noise ratios. 
We report our results using the mean, the statistical uncertainty, and the systematic uncertainty throughout.

The solid line in Fig.~\ref{fig:cb-joint} indicates $\beta=0^{\circ}$, while the dotted line shows $\beta=0.34^{\circ}$, corresponding to the best-fit joint analysis of \Planck\ PR4 and \WMAP\ 9-year result \citep{Eskilt:2022cff}. The dashed lines show the corresponding measurements from the 9-year \WMAP\ and PR4 LFI data sets in each panel. Their statistical error bars are similar to the measurement from \cosmoglobe\ and \BP\ and not shown. We note that the 9-year \WMAP\ gives a similar result to the \cosmoglobe\ \WMAP\ channels, while the \BP\ LFI measurements of $\beta$ are roughly $1\sigma$ higher than the same measurements from \Planck\ PR4 LFI data. This is likely due to the fact that \BP\ fixed the null test failures that plagued the official \planck\ LFI channels.

We conservatively discard the first 10 samples in each Markov chain as burn-in, but note that we have not seen any effect that indicates a significant non-stationary period. Averaging over all post-burn-in samples, we find ${\beta = 0.81^\circ\pm 2.40^\circ(\mathrm{stat.})\pm 0.42^\circ(\mathrm{syst.})}$ for \WMAP; ${\beta = 0.47^\circ \pm 0.93^\circ \pm 0.36^\circ}$ for LFI; and ${\beta=0.35^\circ \pm 0.66^\circ\pm 0.23^\circ}$ for the joint analysis of both LFI and \WMAP. These results are tabulated in Table~\ref{tab:beta}, where the statistical and systematic uncertainties are added in quadrature.

We also show the combined measurements in Fig.~\ref{fig:wmap_lfi_joint_hist}, where the histograms show the scattering of the central values, while the red and blue lines show the measurements with the statistical and total uncertainty, respectively. Here we note that the instrumental systematic uncertainties of $\beta$ are much lower than statistical uncertainties from noise for both \WMAP\ and LFI. Specifically, we find $\sigma_{\mathrm{syst}}/\sigma_{\mathrm{stat}} = 39\%$ for LFI and $\sigma_{\mathrm{syst}}/\sigma_{\mathrm{stat}} = 17\%$ for \WMAP. At the same time, the total uncertainties are too large to shed independent light on the HFI-based result of $\beta \sim 0.35^\circ$.

\begin{figure}
\centering
\includegraphics[width=\linewidth]{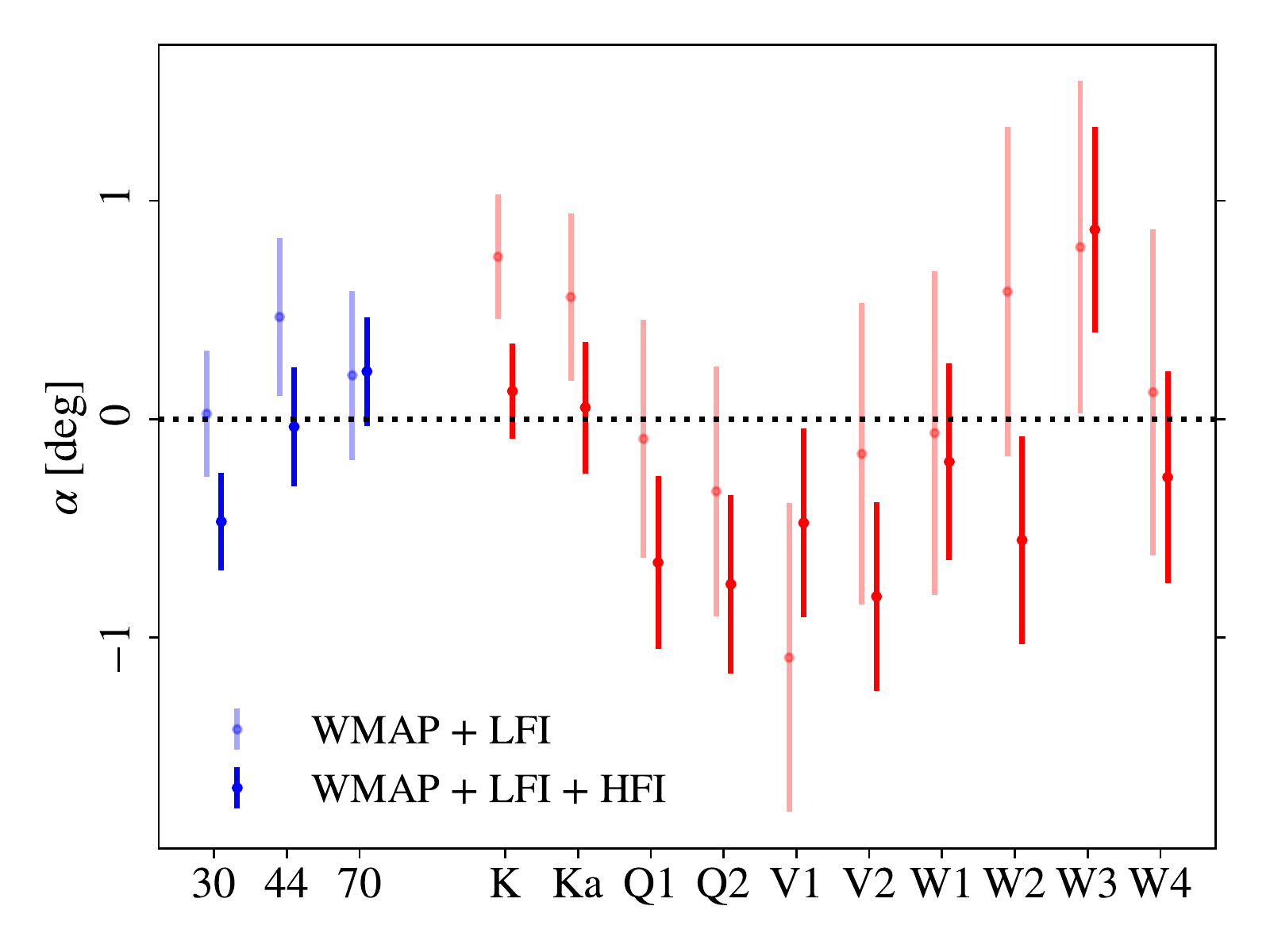}
\caption{Miscalibration angles measured in \WMAP\ (red) and \planck\ LFI (blue). The semi-transparent lines show the measurements equivalent to the lower panels of Figs.~\ref{fig:cb-joint} and \ref{fig:wmap_lfi_joint_hist}, while the opaque lines correspond to the $\beta$ measurement in Fig.~\ref{fig:beta_bp_wmap_lfi_npipe_hfi_dust_vs_synch}. The error bars include both statistical and systematic uncertainties.}
\label{fig:miscalib}
\end{figure}

\begin{table}[t]       %  Table 1
  \begingroup                                                                            
  \newdimen\tblskip \tblskip=5pt
  \caption{Summary of posterior mean miscalibration angles, $\alpha$, for each \Planck\ LFI and \WMAP\ frequency channel, using either only LFI and \WMAP\ data for fitting the model (second column) or additionally HFI data (third column). All uncertainties are 68\,\% errors, and include both statistical and systematic contributions.
    \label{tab:alpha}}
  \nointerlineskip                                                                                                                                                                                     
  \vskip -2mm
  \footnotesize                                                                                                                                      
  \setbox\tablebox=\vbox{ %                                                                                                                                                                             
  \newdimen\digitwidth                                                                                                                          
  \setbox0=\hbox{\rm 0}
  \digitwidth=\wd0
  \catcode`*=\active
  \def*{\kern\digitwidth}
  \newdimen\signwidth
  \setbox0=\hbox{+}
  \signwidth=\wd0
  \catcode`!=\active
  \def!{\kern\signwidth}
  \newdimen\decimalwidth
  \setbox0=\hbox{.}
  \decimalwidth=\wd0
  \catcode`@=\active
  \def@{\kern\signwidth}
  \halign{ \hbox to 1in{#\leaderfil}\tabskip=0.4em&
    \hfil$#$\hfil\tabskip=0.8em&        
    \hfil$#$\hfil\tabskip=0em\cr
    \noalign{\doubleline}
    \omit&\multispan2\hfil $\alpha\, \mathrm{[deg]}$ \hfil\cr
    \noalign{\vskip -3pt}
    \omit&\multispan2\hrulefill\cr
    \noalign{\vskip 3pt}         
  \omit{Channel}\hfil&\omit{LFI+\WMAP}\hfil&\omit\hfil{LFI+\WMAP+HFI} \hfil\cr
  \noalign{\vskip 3pt\hrule\vskip 5pt}
  LFI \quad\,\,30\,GHz& \phantom{-}0.02 \pm 0.29 &-0.47\pm0.22 \cr
  \phantom{LFI} \quad\,\,44\,GHz & \phantom{-}0.47 \pm 0.36           &-0.03\pm0.27 \cr
  \phantom{LFI} \quad\,\,70\,GHz & \phantom{-}0.20 \pm 0.39 &\phantom{-}0.22\pm 0.25 \cr
  \noalign{\vskip 5pt}       
  \WMAP\ \K & \phantom{-}0.74\pm0.29   & \phantom{-}0.13\pm0.22 \cr 
 \phantom{\WMAP} \Ka & \phantom{-}0.56 \pm 0.38 & \phantom{-}0.05 \pm 0.30 \cr 
\phantom{\WMAP} \Q1 & -0.09\pm0.55   & -0.66 \pm 0.40 \cr 
\phantom{\WMAP} \Q2 & -0.33\pm0.57   & -0.76 \pm 0.41 \cr 
\phantom{\WMAP} \V1 & -1.09\pm0.71   & -0.47 \pm 0.43 \cr 
\phantom{\WMAP} \V2 & -0.16\pm0.69   & -0.81 \pm 0.43 \cr 
\phantom{\WMAP} \W1 & -0.06\pm0.74   & -0.20 \pm 0.45 \cr 
\phantom{\WMAP} \W2 & \phantom{-}0.58\pm0.75   & -0.55 \pm 0.48 \cr 
\phantom{\WMAP} \W3 & \phantom{-}0.79\pm0.76   & \phantom{-}0.87 \pm 0.47 \cr 
\phantom{\WMAP} \W4 & \phantom{-}0.12\pm0.75   & -0.27 \pm 0.49 \cr 
    \noalign{\vskip 3pt\hrule\vskip 5pt}   
  }}
  \endPlancktablewide                                                                                                                                            
  \endgroup
\end{table}

The corresponding miscalibration angles, $\alpha$, are tabulated in Table~\ref{tab:alpha} and plotted in Fig.~\ref{fig:miscalib}. Overall, we see that these are generally consistent with zero, with the \WMAP\ \K-band showing the largest deviation at $2.6\sigma$ for the \WMAP+LFI-only analysis. However, this result is not robust with respect to data selection.

\subsection{LFI+HFI analysis}

\begin{figure}
\centering
\includegraphics[width=\linewidth]{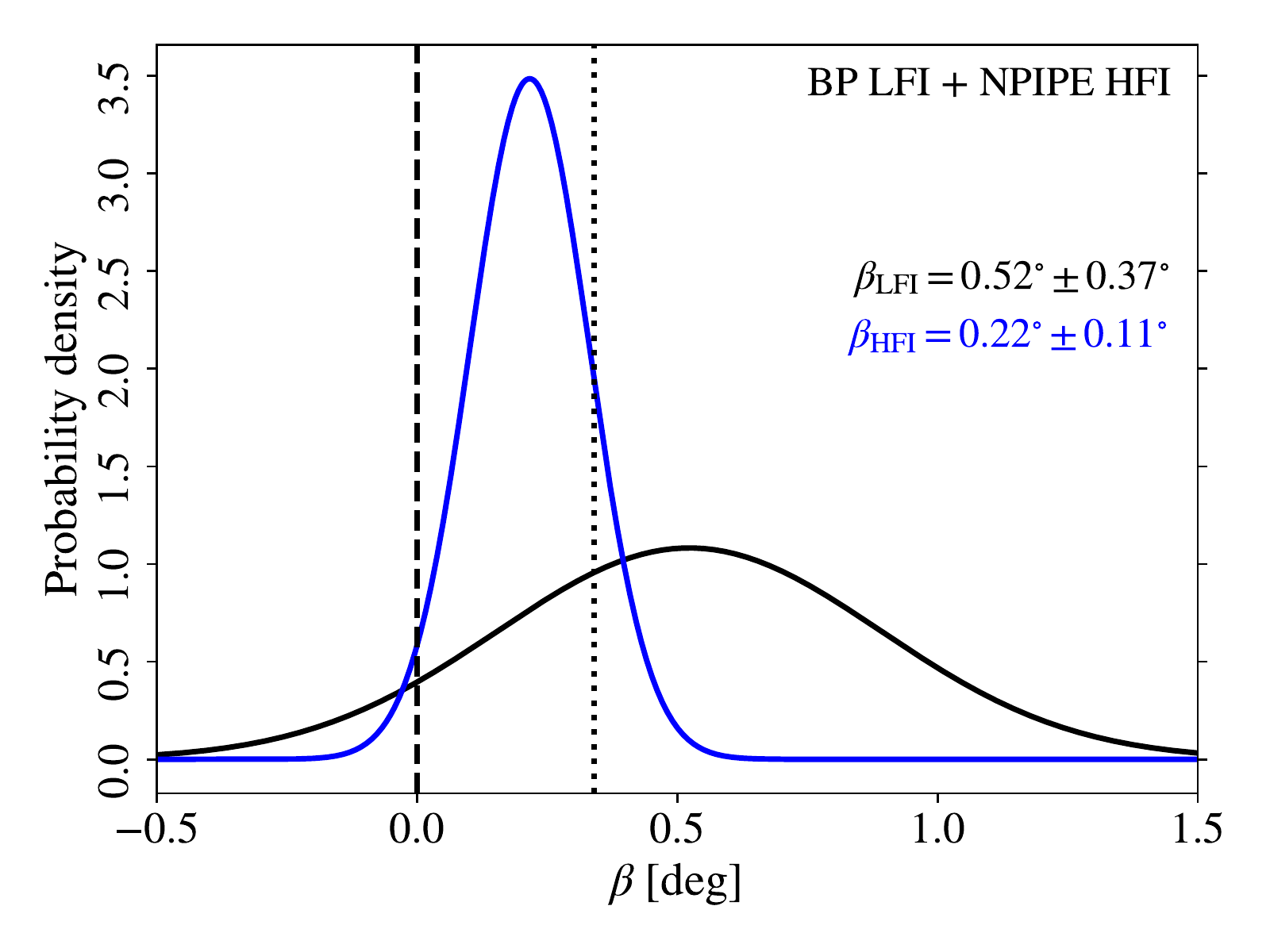}
\caption{Probability density of cosmic birefringence for every fifth Gibbs iteration of the \BP{} LFI maps. This is analyzed jointly with the \Planck\ PR4 HFI detector split maps. One birefringence angle is additionally sampled for the LFI channels and another for the HFI channels. The error bars include both statistical and systematic uncertainties from LFI. The dashed vertical line shows $\beta = 0^\circ$ while the dotted line shows $\beta = 0.34^\circ$.}
\label{fig:beta_bp_lfi_npipe_hfi}
\end{figure}

Next, we perform a joint analysis between the \BP\ LFI and \Planck\ PR4 HFI maps. In this analysis, we include the dust filamentary model for HFI, and we sample jointly two cosmic birefringence angles $\beta_i$, eleven miscalibration angles $\alpha_i$, and four dust amplitude parameters $A_\ell$. Since we are including more data and the dust filamentary model for dust $EB$ is computationally more expensive, we sample $\beta$ for every fifth Gibbs iteration. Fig.~\ref{fig:beta_bp_lfi_npipe_hfi} shows the resulting constraints on $\beta$ separately for the LFI and the HFI channels.

Note that we have only access to one set of maximum-likelihood \Planck\ PR4 HFI maps, and it is therefore harder to assess the impact of instrumental systematic effects than it is with the \BP\ and \cosmoglobe\ maps. However, \cite{Diego-Palazuelos:2022cnh} considered a wide range of effects, and found these to have a limited impact on $\beta$.

\begin{figure}
\centering
\includegraphics[width=\linewidth]{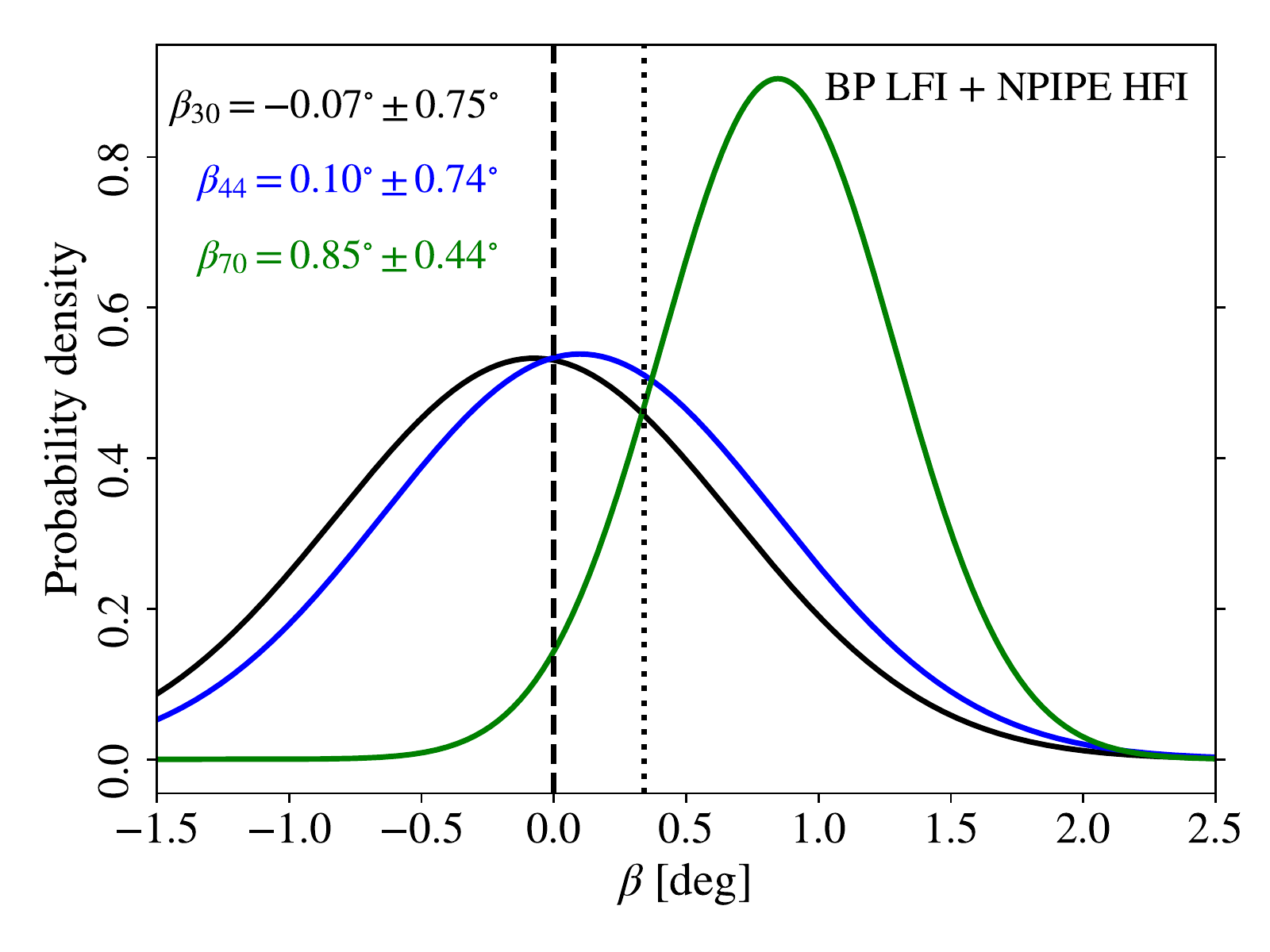}
\caption{Same as Fig.~\ref{fig:beta_bp_lfi_npipe_hfi}, except one $\beta$ is sampled individually for the 30, 44, and 70\, GHz channels. One $\beta$ is additionally sampled for HFI but not shown. We show the total uncertainties from both LFI systematics and statistical uncertainties.}
\label{fig:beta_bp_lfi_npipe_hfi_lfi_ind}
\end{figure}

We find ${\beta_{\textrm{LFI}} = 0.52^\circ \pm 0.34^\circ\pm 0.13^\circ}$ which gives a combined uncertainty of $0.37^\circ$. This corresponds nominally to a non-zero birefringence angle of $1.4\sigma$. To understand which channels drive this value, we fit $\beta$ individually for each LFI channel. This is shown in terms of posterior distributions in Fig.~\ref{fig:beta_bp_lfi_npipe_hfi_lfi_ind}, and posterior summary statistics are tabulated in Table~\ref{tab:beta}. As usual, we additionally sample a separate birefringence angle for HFI $\beta_{\mathrm{HFI}}$, miscalibration angles $\alpha_i$, and dust $EB$ amplitudes $A_\ell$ which are not shown.

In Fig.~\ref{fig:beta_bp_lfi_npipe_hfi_lfi_ind} we see that the positive value is dominated by the 70\,GHz channel, with ${\beta_{70} = 0.85^{\circ}\pm0.42^\circ\pm0.13^\circ}$, which corresponds to a total uncertainty of $0.44^\circ$. This formally disfavors $\beta=0^\circ$ at a statistical significance of $1.9\sigma$. In contrast, the 30 and 44\,GHz channels show no evidence for a positive value, but their uncertainties are also significantly larger, with posterior summary values of ${\beta_{30} = -0.07^\circ \pm 0.69^\circ\pm0.29^\circ}$ and ${\beta_{44} = 0.10^\circ \pm 0.70^{\circ}\pm0.25^\circ}$, respectively. We also note that, while $\beta_{44}$ and $\beta_{70}$ are consistent with the \Planck\ PR4 LFI analysis described by \cite{Eskilt:2022wav}, which used a similar mask, the 30\,GHz channel measurement from \BP\ data has dropped by more than $1\sigma$ as compared to the \Planck\ PR4 LFI maps.

One might wonder if the large measurement of $\beta$ in the 70\,GHz channel could be caused by polarized dust $EB$ emission. This channel contains more polarized synchrotron than dust emission, but the latter is not negligible as the channel is close to the foreground minimum where synchrotron and dust emission are roughly equal \citep{planck2016-l04}. We have not applied our dust filamentary model to this channel, but we note that from earlier work the birefringence angle always increases when accounting for this effect \citep{Diego-Palazuelos:2022dsq}, and therefore the quoted value of $\beta_{70}$ represents a conservative lower limit. 

For the HFI channels, we find ${\beta_{\textrm{HFI}}=0.22^\circ\pm0.11^\circ\pm 0.02^\circ}$, where the systematic uncertainty indicates only the variation due to LFI marginalization. This value is slightly lower than previously reported values, and parts of this is likely due to the somewhat more conservative mask used in the current analysis, due to missing pixels in the half-mission split \citep{Diego-Palazuelos:2022dsq, Eskilt:2022wav, Eskilt:2022cff}. We also find that the $C^{TB}_\ell / C^{TE}_\ell$ power ratio is higher in the 353\,GHz channel at smaller multipoles for the mask used in this work. This is a tracer of polarized dust emission that is known to be highly mask-dependent \citep{Clark:2021kze, Diego-Palazuelos:2022dsq}, and this might suggest a larger dust $EB$ amplitude that might not be fully encapsulated by the dust model. \cite{Diego-Palazuelos:2022dsq} showed that the dust filamentary model is mostly able to mitigate the effect of dust, but one would still expect a slight weakening of the signal of $\beta$ for masks with a sky coverage around $f_{\textrm{sky}} \lesssim 0.90$.

\begin{figure}
\centering
\includegraphics[width=\linewidth]{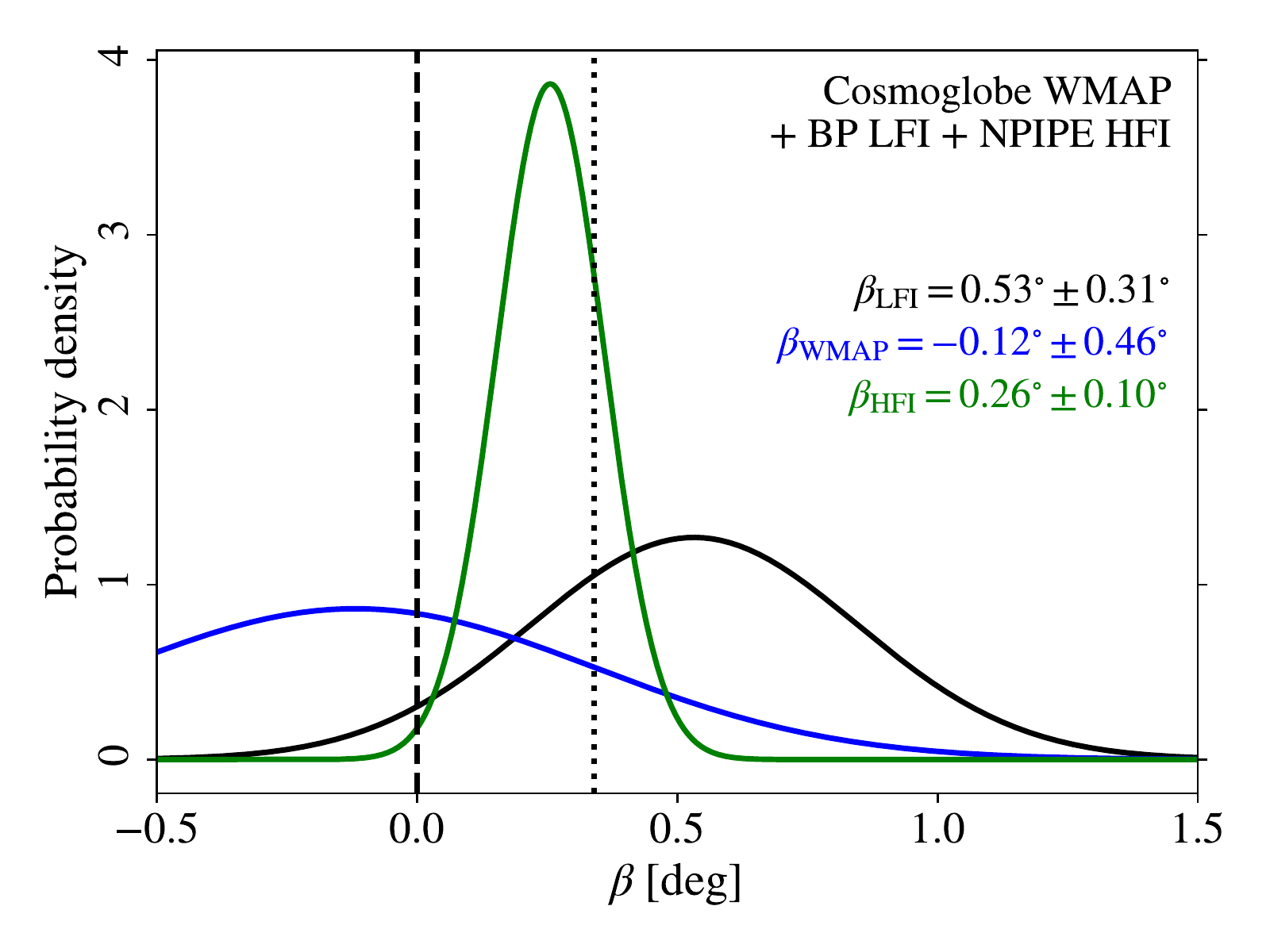}
\caption{Same as Fig.~\ref{fig:beta_bp_lfi_npipe_hfi} but jointly with the \cosmoglobe\ \WMAP\ channels for every 16th Gibbs iteration. An additional $\beta$ is sampled for the \WMAP\ channels.}
\label{fig:beta_bp_wmap_lfi_npipe_hfi}
\end{figure}

\subsection{\WMAP+LFI+HFI analysis}

Finally, we analyze the combination of all three main data sets, namely \cosmoglobe\ 
\WMAP, \BP\ LFI, and \Planck\ PR4 HFI channels. The main results from this analysis are summarized in Fig.~\ref{fig:beta_bp_wmap_lfi_npipe_hfi}, where $\beta$ is measured separately for \WMAP, LFI, and HFI. Only every 16th Gibbs sample is processed in this analysis, due to a high computational cost.

For \WMAP, we find that ${\beta_{\mathit{WMAP}} = -0.12^\circ \pm 0.46^\circ\pm0.06^\circ}$, where the systematic error term denotes the systematic uncertainties arising only from LFI and \WMAP, and not HFI. For LFI, we find ${\beta_{\textrm{LFI}} = 0.53^\circ \pm 0.31^\circ\pm0.08^\circ}$, which is inconsistent with ${\beta=0^\circ}$ at a $1.6\sigma$ statistical significance. Finally, for HFI we find ${\beta_{\textrm{HFI}}= 0.26^\circ \pm 0.10^\circ\pm0.01^\circ}$. We note that even though the central value of $\beta_{\mathit{WMAP}}$ itself is negative, the addition of the \WMAP\ channels slightly increases the statistical significance of both $\beta_{\mathrm{LFI}}$ and $\beta_{\mathrm{HFI}}$ by providing more information through cross-correlations.

\begin{figure}
\centering
\includegraphics[width=\linewidth]{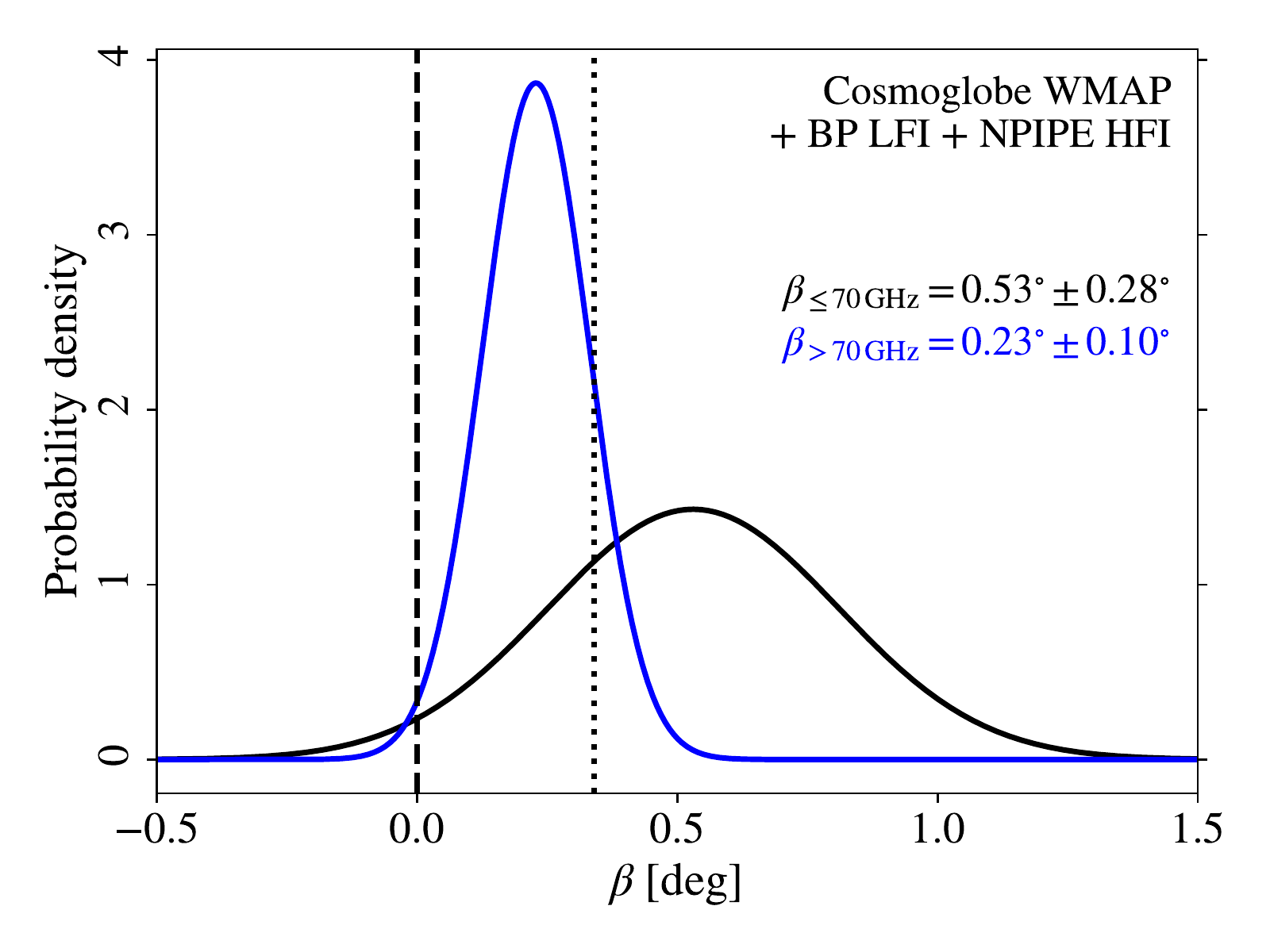}
\caption{Similar to Fig.~\ref{fig:beta_bp_wmap_lfi_npipe_hfi} but we sample $\beta$ separately for synchrotron-dominated channels, $\beta_{\leq 70\,\textrm{GHz}}$, and dust-dominated channels, $\beta_{> 70\,\textrm{GHz}}$.}
\label{fig:beta_bp_wmap_lfi_npipe_hfi_dust_vs_synch}
\end{figure}

To further probe the impact of channels that are dominated by either synchrotron or thermal dust emission, we split $\beta$ into two groups; one for synchrotron-dominated channels, $\beta_{\leq 70\,\textrm{GHz}}$, and another for thermal dust-dominated channels, $\beta_{> 70\,\textrm{GHz}}$. The results from this calculation are summarized in Fig.~\ref{fig:beta_bp_wmap_lfi_npipe_hfi_dust_vs_synch} and Table~\ref{tab:beta}.

For the synchrotron-dominated channels, we find ${\beta_{\leq 70\,\mathrm{GHz}} = 0.53^{\circ}\pm0.27^\circ\pm0.06^\circ}$ which disfavors $\beta=0^{\circ}$ at a $1.9\sigma$ statistical significance. This signal is mostly driven by the 70\,GHz channel, but as noted above, the addition of \WMAP\ does increase the statistical significance of $\beta_{\textrm{LFI}}$ which, with the exception of the \W-band, belong to the channels with frequency $\leq 70\,\mathrm{GHz}$. 

While the dust filamentary model of \citet{Clark:2021kze} is not applicable to synchrotron emission, it is possible that a different mechanism gives rise to an intrinsic synchrotron $EB$ power spectrum. Negative synchrotron $EB$ correlations would bias $\beta$ to be larger and could be the cause of the signal. Another possibility is $EB$ correlations between polarized synchrotron and dust emission. Knowing that synchrotron and dust emission are correlated \citep{Choi:2015xha, planck2016-l11A}, we can not rule out the existence of $C^{E_{\mathrm{dust}}B_{\mathrm{synch}}}_\ell$ or $C^{E_{\mathrm{synch}}B_{\mathrm{dust}}}_\ell$. Lastly, the dust-dominated channels yield ${\beta_{> 70\,\mathrm{GHz}} = 0.23^{\circ}\pm0.10^\circ\pm0.01^\circ}$.

Finally, we compute one global $\beta$ angle for the joint analysis of all three data sets. These results are listed in Table~\ref{tab:beta}, and compared with those found in earlier analyses \citep{Diego-Palazuelos:2022dsq, Eskilt:2022cff}. Joint analysis of $\beta$ is driven by the HFI channels so the signal is sensitive to dust $EB$ which is modelled with the dust filamentary ansatz. The discrepancy of $\beta$ between the left and right columns of the lower half of Table~\ref{tab:beta} is most likely due to the masking sensitivity of the filamentary dust model.

\section{Conclusions}
\label{sec:conclusions}

We have analyzed the reprocessed \BP\ LFI and \cosmoglobe\ \WMAP\ polarization maps with respect to isotropic cosmic birefringence. These channels shed new light on the previously reported measurement of a non-zero birefringence angle $\beta$ from \Planck\ HFI, as they both have very different systematics from HFI, and they are dominated by synchrotron emission, for which no significant $EB$ correlations have been measured \citep{Martire:2021gbc, Rubino-Martin:2023fya}. While there has been no direct measurement of the $EB$ correlations of thermal dust emission, there are indications for its existence \citep{Huffenberger:2019mjx, Clark:2021kze}.

The uncertainty on $\beta$ from LFI and \WMAP\ alone is too large to allow for robust independent evidence, with a best-fit combined value of ${\beta = 0.35^\circ \pm 0.70^\circ}$, where the error bar includes both systematic and statistical uncertainties. This uncertainty is more than six times larger than that for HFI alone. However, when performing a joint analysis with \Planck\ PR4 HFI, which provides additional valuable information regarding the foregrounds and miscalibration angles, we do find a value of $\beta = 0.53^\circ\pm0.28^\circ$ in the synchrotron-dominated maps which include channels up to and including 70\,GHz. This is positive at a statistical significance of $1.9\sigma$ and is consistent with $\beta = 0.342_{\phantom{\circ}-0.091^{\circ}}^{\circ + 0.094^\circ}$ from the joint analysis of the public 9-year \WMAP\ and \Planck\ PR4 data \citep{Eskilt:2022cff}.

The measurement from the synchrotron-dominated channels is driven by the 70\,GHz channel, which is notable for at least two reasons. Firstly, this channel has the lowest foreground contribution of any channel considered in this line of work. Secondly, it is the LFI channel with the highest sensitivity and lowest instrumental systematics, which is why it has formed the basis of almost all cosmological results from LFI. 

Although there has been found no evidence for synchrotron $EB$ correlations to date, we can neither rule out its existence nor possible correlations between $E$ and $B$ modes of synchrotron and dust emission, namely $C^{E_{\mathrm{dust}}B_{\mathrm{synch}}}_{\ell}$ and $C^{E_{\mathrm{synch}}B_{\mathrm{dust}}}_{\ell}$. This could bias our measurements since we measure birefringence in synchrotron channels also through cross-power spectra with the dust-dominated HFI channels.

While the current analysis suggests that the current instrumental model adopted in the \cosmoglobe\ framework does not play a dominant role regarding the final uncertainties, accounting for typically only about 10\,\% of the total uncertainty, it is important to emphasize that the polarization angle is not yet sampled over directly in this framework. Adding support for that degree of freedom could be important in order to decrease these uncertainties further, for instance using Tau~A as a calibration source at the map level. This should indeed be considered a high-priority issue for next-generation processing for both \Planck\ and \WMAP.

Future CMB experiments should, however, focus their efforts on precise on-ground calibrations of their polarization angles, which is no longer possible for \WMAP\ and \Planck. Negligible miscalibration angles make this analysis more robust against foreground $EB$ as we do not have to break the $\alpha+\beta$ degeneracy. See \citet{Cornelison:2022zrc} where the authors achieved a polarization angle uncertainty of $<0.1^\circ$ for BICEP3.

If cosmic birefringence is demonstrated to be of cosmological origin, it would be a revolutionary window on high-energy physics and cosmology \citep{Agrawal:2022lsp, Fujita:2020ecn, Komatsu:2022nvu}, and possibly shed light on the dark sector of the Universe.

\begin{acknowledgements}
    We are grateful to Patricia Diego-Palazuelos and Eiichiro Komatsu for useful discussions and feedback. \planck\ is a project of the European Space Agency (ESA) with instruments provided by two scientific consortia funded by ESA member states and led by Principal Investigators from France and Italy, telescope reflectors provided through a collaboration between ESA and a scientific consortium led and funded by Denmark, and additional contributions from NASA (USA). The current work has received funding from the European Union’s Horizon 2020 research and innovation programme under grant agreement numbers 819478 (ERC; Cosmoglobe)
    and 772253 (ERC; bits2cosmology). In addition, the collaboration acknowledges support from RCN (Norway; grant no. 274990). Softwares: PolSpice \citep{Chon:2003gx}, healpy \citep{Gorski:2004by, Zonca2019}, Matplotlib \citep{Hunter:2007}, NumPy \citep{2020NumPy-Array}, CAMB \citep{Lewis:2000}, emcee \citep{ForemanMackey:2012ig}.
\end{acknowledgements}

\bibliographystyle{aa}
\bibliography{references}

\end{document}

%% file: Planck.tex
\def\setsymbol#1#2{\expandafter\def\csname #1\endcsname{#2}}
\def\getsymbol#1{\csname #1\endcsname}

\def\Planck{\textit{Planck}}

\newbox\tablebox    \newdimen\tablewidth
\def\leaderfil{\leaders\hbox to 5pt{\hss.\hss}\hfil}
\def\endPlancktablewide{\tablewidth=\textwidth 
    $$\hss\copy\tablebox\hss$$
    \vskip-\lastskip\vskip -2pt}
\def\tablenote#1 #2\par{\begingroup \parindent=0.8em
    \abovedisplayshortskip=0pt\belowdisplayshortskip=0pt
    \noindent
    $$\hss\vbox{\hsize\tablewidth \hangindent=\parindent \hangafter=1 \noindent
    \hbox to \parindent{$^#1$\hss}\strut#2\strut\par}\hss$$
    \endgroup}
\def\doubleline{\vskip 3pt\hrule \vskip 1.5pt \hrule \vskip 5pt}

\def\L2{\ifmmode L_2\else $L_2$\fi}

\def\DeltaT{\ifmmode \Delta T\else $\Delta T$\fi}
\def\deltat{\ifmmode \Delta t\else $\Delta t$\fi}
\def\fknee{\ifmmode f_{\rm knee}\else $f_{\rm knee}$\fi}
\def\Fmax{\ifmmode F_{\rm max}\else $F_{\rm max}$\fi}
\def\solar{\ifmmode{\rm M}_{\mathord\odot}\else${\rm M}_{\mathord\odot}$\fi}
\def\Msolar{\ifmmode{\rm M}_{\mathord\odot}\else${\rm M}_{\mathord\odot}$\fi}
\def\Lsolar{\ifmmode{\rm L}_{\mathord\odot}\else${\rm L}_{\mathord\odot}$\fi}
\def\inv{\ifmmode^{-1}\else$^{-1}$\fi}
\def\mo{\ifmmode^{-1}\else$^{-1}$\fi}
\def\sup#1{\ifmmode ^{\rm #1}\else $^{\rm #1}$\fi}
\def\expo#1{\ifmmode \times 10^{#1}\else $\times 10^{#1}$\fi}
\def\,{\thinspace}
\def\lsim{\mathrel{\raise .4ex\hbox{\rlap{$<$}\lower 1.2ex\hbox{$\sim$}}}}
\def\gsim{\mathrel{\raise .4ex\hbox{\rlap{$>$}\lower 1.2ex\hbox{$\sim$}}}}

\def\simprop{\mathrel{\raise .4ex\hbox{\rlap{$\propto$}\lower 1.2ex\hbox{$\sim$}}}}
\def\deg{\ifmmode^\circ\else$^\circ$\fi}
\def\pdeg{\ifmmode $\setbox0=\hbox{$^{\circ}$}\rlap{\hskip.11\wd0 .}$^{\circ}
          \else \setbox0=\hbox{$^{\circ}$}\rlap{\hskip.11\wd0 .}$^{\circ}$\fi}
\def\arcs{\ifmmode {^{\scriptstyle\prime\prime}}
          \else $^{\scriptstyle\prime\prime}$\fi}
\def\arcm{\ifmmode {^{\scriptstyle\prime}}
          \else $^{\scriptstyle\prime}$\fi}
\newdimen\sa  \newdimen\sb
\def\parcs{\sa=.07em \sb=.03em
     \ifmmode \hbox{\rlap{.}}^{\scriptstyle\prime\kern -\sb\prime}\hbox{\kern -\sa}
     \else \rlap{.}$^{\scriptstyle\prime\kern -\sb\prime}$\kern -\sa\fi}
\def\parcm{\sa=.08em \sb=.03em
     \ifmmode \hbox{\rlap{.}\kern\sa}^{\scriptstyle\prime}\hbox{\kern-\sb}
     \else \rlap{.}\kern\sa$^{\scriptstyle\prime}$\kern-\sb\fi}
\def\ra[#1 #2 #3.#4]{#1\sup{h}#2\sup{m}#3\sup{s}\llap.#4}
\def\dec[#1 #2 #3.#4]{#1\deg#2\arcm#3\arcs\llap.#4}
\def\deco[#1 #2 #3]{#1\deg#2\arcm#3\arcs}
\def\rra[#1 #2]{#1\sup{h}#2\sup{m}}

\def\dots{\relax\ifmmode \ldots\else $\ldots$\fi}
\def\WHzsr{\ifmmode $W\,Hz\mo\,sr\mo$\else W\,Hz\mo\,sr\mo\fi}
\def\mHz{\ifmmode $\,mHz$\else \,mHz\fi}
\def\GHz{\ifmmode $\,GHz$\else \,GHz\fi}
\def\mKs{\ifmmode $\,mK\,s$^{1/2}\else \,mK\,s$^{1/2}$\fi}
\def\muKs{\ifmmode \,\mu$K\,s$^{1/2}\else \,$\mu$K\,s$^{1/2}$\fi}
\def\muKRJs{\ifmmode \,\mu$K$_{\rm RJ}$\,s$^{1/2}\else \,$\mu$K$_{\rm RJ}$\,s$^{1/2}$\fi}
\def\muKHz{\ifmmode \,\mu$K\,Hz$^{-1/2}\else \,$\mu$K\,Hz$^{-1/2}$\fi}
\def\MJysr{\ifmmode \,$MJy\,sr\mo$\else \,MJy\,sr\mo\fi}
\def\MJysrmK{\ifmmode \,$MJy\,sr\mo$\,mK$_{\rm CMB}\mo\else \,MJy\,sr\mo\,mK$_{\rm CMB}\mo$\fi}
\def\microns{\ifmmode \,\mu$m$\else \,$\mu$m\fi}

\def\muK{\ifmmode \,\mu$K$\else \,$\mu$\hbox{K}\fi}
\def\microK{\ifmmode \,\mu$K$\else \,$\mu$\hbox{K}\fi}
\def\muW{\ifmmode \,\mu$W$\else \,$\mu$\hbox{W}\fi}
\def\kms{\ifmmode $\,km\,s$^{-1}\else \,km\,s$^{-1}$\fi}
\def\kmsMpc{\ifmmode $\,\kms\,Mpc\mo$\else \,\kms\,Mpc\mo\fi}
\providecommand{\sorthelp}[1]{}

%% file: authors.tex
\newcommand{\oslo}[0]{1}
\newcommand{\iiabangalore}[0]{2}

\author{\small
J.~R.~Eskilt\inst{\ref{uio},\ref{imperial}}\thanks{Corresponding author: J.~R.~Eskilt; \url{j.r.eskilt@astro.uio.no}}
\and
D.~J.~Watts\inst{\ref{uio}}
\and
R.~Aurlien\inst{\ref{uio}}
\and
A.~Basyrov\inst{\ref{uio}}
\and
M.~Bersanelli\inst{\ref{milan}}
\and
M.~Brilenkov\inst{\ref{uio}}
\and
L.~P.~L.~Colombo\inst{\ref{milan}}
\and
H.~K.~Eriksen\inst{\ref{uio}}
\and
K. S. F. Fornazier\inst{\ref{ifusp}}
\and
C.~Franceschet\inst{\ref{milan}}
\and
U.~Fuskeland\inst{\ref{uio}}
\and
M.~Galloway\inst{\ref{uio}}
\and
E.~Gjerl\o w\inst{\ref{uio}}
\and
B.~Hensley\inst{\ref{princeton}}
\and
L.~T.~Hergt\inst{\ref{ubc}}
\and
D.~Herman\inst{\ref{uio}}
\and
H.~T.~Ihle\inst{\ref{uio}}
\and
K.~Lee\inst{\ref{uio}}
\and
J.~G.~S.~Lunde\inst{\ref{uio}}
\and
S.~K.~Nerval\inst{\ref{dunlap1},\ref{dunlap2}}
\and
S.~Paradiso\inst{\ref{waterloo}}
\and
S.~K.~Patel\inst{\ref{iit_bhu}}
\and
F.~Rahman\inst{\ref{iiabangalore}}
\and
M.~Regnier\inst{\ref{apc}}
\and
M.~San\inst{\ref{uio}}
\and
S.~Sanyal\inst{\ref{iit_bhu}}
\and
N.-O.~Stutzer\inst{\ref{uio}}
\and
H.~Thommesen\inst{\ref{uio}}
\and
A.~Verma\inst{\ref{iit_bhu}}
\and
I.~K.~Wehus\inst{\ref{uio}}
\and
Y.~Zhou\inst{\ref{berkeley}}
}
\institute{\small
Institute of Theoretical Astrophysics, University of Oslo, Blindern, Oslo, Norway\label{uio}
\and
Imperial Centre for Inference and Cosmology, Department of Physics, Imperial College London, Blackett Laboratory, Prince Consort Road, London SW7 2AZ, United Kingdom\label{imperial}
\and
Dipartimento di Fisica, Università degli Studi di Milano, Via Celoria, 16, Milano, Italy\label{milan}
\and
Instituto de Física, Universidade de São Paulo - C.P. 66318, CEP: 05315-970, São Paulo, Brazil\label{ifusp}
\and
Department of Astrophysical Sciences, Princeton University, 4 Ivy Lane, Princeton, NJ 08540\label{princeton}
\and
Department of Physics and Astronomy, University of British Columbia, 6224 Agricultural Road, Vancouver BC, V6T1Z1, Canada\label{ubc}
\and
David A. Dunlap Department of Astronomy \& Astrophysics, University of Toronto, 50 St. George Street, Toronto, ON M5S 3H4, Canada\label{dunlap1}
\and
Dunlap Institute for Astronomy \& Astrophysics, University of Toronto, 50 St. George Street, Toronto, ON M5S 3H4, Canada\label{dunlap2}
\and
Waterloo Centre for Astrophysics, University of Waterloo, Waterloo, ON N2L 3G1, Canada\label{waterloo}
\and
Department of Physics, Indian Institute of Technology (BHU), Varanasi - 221005, India\label{iit_bhu}
\and
Indian Institute of Astrophysics, Koramangala II Block, Bangalore, 560034, India\label{iiabangalore}
\and
Laboratoire Astroparticule et Cosmologie (APC), Université Paris-Cité, Paris, France\label{apc}
\and
Department of Physics, University of California, Berkeley, Berkeley, CA 94720, USA\label{berkeley}
}

 %\author{V.~Arsenijevic\inst{\ref{inst1}}\and S.~Fabbro\inst{\ref{inst2}}\and
%A.~M.~Mour\~ao\inst{\ref{inst3}}\and A.~J.~Rica da Silva\inst{\ref{inst1}}}
%
%\institute{Multidisciplinar de Astrof\'{\i}sica, IST, Avenida Rovisco Pais, 1049
%Lisbon, Portugal\email{...}\label{inst1} \and < Multidisciplinar de Astrof\'{\i}sica, IST, Avenida Rovisco Pais, 1049 Lisbon, Portugal\email{...}\label{inst2}
%\and
%Multidisciplinar de Astrof\'{\i}sica, IST, Avenida Rovisco Pais, 1049
%Lisbon, Portugal\email{...}\label{inst3}
%} 